\documentclass[12pt,a4paper]{article}
\usepackage{blindtext}
\usepackage{graphicx}				
\usepackage{packages}
\usepackage{authblk}		
\usepackage{cite}
\usepackage{amssymb}
\usepackage{amssymb,amsmath,amsthm,booktabs,mathtools}
\usepackage{xcolor}
\usepackage{soul}

\newcommand{\p}{\partial}

\newcommand{\dd}{{\rm d}}

\newcommand{\dr}{\mathrm{dr}}

\newcommand{\n}{\nonumber\\}
\newcommand{\ii}{{\rm i}}

\newcommand{\tot}{{\rm tot}}

\newcommand{\eff}{{\rm eff}}
\newcommand{\beq}{\begin{eqnarray}}
\newcommand{\eeq}{\end{eqnarray}}

\numberwithin{equation}{section}
\begin{document}
\title{Form factors and generalized hydrodynamics for integrable systems}
\author[1]{Axel Cort\'es Cubero }
\affil[1]{\small College of Business, University of Puerto Rico at Mayag\"{u}ez, PR-108,  00682 Mayag\"{u}ez, Puerto Rico.}
\author{Takato Yoshimura}
\affil{\small Department of Physics, Tokyo Institute of Technology, Ookayama 2-12-1, Tokyo 152-8551, Japan.}
\author{Herbert Spohn}
\affil{\small Departments of Mathematics and Physics, Technical University Munich,Boltzmannstr. 3, 85747 Garching, Germany.}
\maketitle
\begin{abstract}
Our review covers microscopic foundations of generalized hydrodynamics (GHD). As one generic approach we develop form factor expansions, 
for ground states and generalized Gibbs ensembles (GGE).  In the latter case the so obtained results are compared with predictions from GHD. 
One cornerstone of GHD are the GGE averaged microscopic currents, which can also be obtained through employing form factors. 
Discussed is a second, completely orthogonal approach based on the availability of a self-conserved current.
\end{abstract}

\tableofcontents

\section{Introduction}
Hydrodynamics is one of the oldest theoretical branches in physics, providing the mathematical equations governing the motion of fluids. 
A fluid possesses five locally conserved fields in three dimensions and, as a consequence, the Euler equations are a system of five coupled hyperbolic 
conservation laws. Their validity is based on the physically very natural hypothesis of the propagation of local equilibrium.
Since then the hydrodynamic method has been applied in many areas, from Bose-Einstein condensates, to magneto-hydrodynamics,
 and to the formation of super-dense neutron stars, to mention only a few. 
 
 A much more recent addition is the hydrodynamics of  integrable many-body systems.
 As novel feature, the number of conserved fields is extensive and, in the large scale limit, one has to deal with an infinite number of coupled conservation laws. 
The hydrodynamic theory of integrable systems goes by the name of {\it generalized hydrodynamics} (GHD) \cite{PhysRevX.6.041065,PhysRevLett.117.207201}. GHD has been successfully applied to a number of integrable systems, both quantum \cite{PhysRevB.96.020403,PhysRevLett.119.020602,PhysRevLett.119.220604,PhysRevLett.120.045301,PhysRevLett.119.195301,DOYON2018570,PhysRevB.96.115124,PhysRevB.97.081111,SciPostPhys.2.2.014,PhysRevLett.121.160603,PhysRevLett.124.140603} and classical \cite{10.21468/SciPostPhys.4.6.045,doi:10.1063/1.5096892}, see also \cite{10.21468/SciPostPhysLectNotes.18} for an introductory text. Theoretical predictions based on GHD have also been checked to match with experiments \cite{PhysRevLett.122.090601,malvania2020generalized}. 

To write down the appropriate closed system of hydrodynamic equations, one has to figure out the dependence of the 
macroscopic currents  on the conserved fields. By the assumption of local equilibrium, required are thus the microscopic currents averaged over a Gibbs ensemble. For a simple fluid this is a textbook exercise, the result being expressed in terms of the thermodynamic pressure. For integrable systems the appropriate Gibbs ensemble depends on an extensive number of chemical potentials, therefore called 
\textit{generalized Gibbs ensemble} (GGE) \cite{PhysRevLett.98.050405}. The average currents are no longer simply related to the free energy and the computation of the 
required GGE average is a difficult task. The initial derivation \cite{PhysRevX.6.041065,PhysRevLett.117.207201} is based on a \textit{rate assumption}, as will be explained in more detail. The determining equation involves the two-particle scattering shift. Usually this information  is available from the microscopic model. Thus one had accomplished a scheme which in principle applies to the entire class of integrable many-body systems. Still, from a theoretical perspective, one would like to establish the rate assumption on the basis of a specific microscopic model. 

As part of our article, we will review some attempts for computing the GGE averaged currents. The methods employed involve
the boost operator and the symmetry of the GGE correlator for conserved fields and currents.  A more generic scheme involves thermodynamic \textit{form factor expansions}, which will be discussed extensively.
They can be used not only for the static currents, but also for computing various other correlations. Of particular interest are time-dependent
correlations and also correlations for inhomogeneous states of GGE type. For a homogeneous system, time correlations can be computed also on the basis of linearized hydrodynamic equations. By their derivation, this should properly describe the correlations on ballistic 
space-time scales. Thus it is very instructive to compare with form factor results and to understand whether corrections to
the hydrodynamic  predictions are accessible. In a companion review in the same volume \cite{borsi2021current} discussed is a complementary approach, which handles diagonal matrix elements of the current operator.

This review is divided in two parts: we first briefly introduce the notion of GHD to build up some physical intuition behind such a construction. Brief accounts on form factors in integrable systems, including relativistic integrable quantum field theories, quantum spin chains, and the Lieb-Liniger model then follow. Having such techniques at our disposal, we move on to explain the notion of thermodynamic form factors and discuss how some of the predictions from GHD can be reproduced by the thermodynamic form factor approach. The second part of the review deals with GGE averaged currents. In particular, we discuss the validation  of the rate ansatz and compare the power of various approaches, including form factors.

\section{Generalized hydrodynamics}\label{ghdintro}
An integrable system can be characterized by having an infinite number of local conservation laws, which somewhat symbolically 
are of the form
\begin{equation}\label{conseq}
     \partial_t q_i(x,t)+\partial_x j_i(x,t)=0, \quad i = 0,1,...\,.
 \end{equation}
Charge densities $q_i(x) = q_i(x,t=0)$ and current densities $j_i(x) = j_i(x,t=0) $ are local operators, resp. local functions on phase space, relative to location $x$. In concrete models, instead of positive integers  it might be more convenient to use more intricate labelling schemes.
Time $t$ refers to the hamiltonian time evolution.

According to statistical mechanics the natural class of time-stationary states of such systems are the normalized Gibbs density matrices
 \begin{equation}\label{gge1}
\varrho=\frac{1}{Z} \exp \big[-\sum_{i=0}^\infty\beta^iQ_i\big].
\end{equation}
Here $Q_i = \int_\mathbb{R}\dd x \,q_i(x)$ and $\{Q_i\}$ is the complete list of locally conserved charges of the system under consideration (the first few of which are usually chosen as particle number $Q_0=N$, total momentum $Q_1=P$, and energy $Q_2=H$) and the $\beta^i$'s are the associated chemical potentials. Eq. \eqref{gge1} has to be considered first in a
finite volume, say $[-L/2,L/2]$, and $Z$ is the corresponding normalizing partition function. 
Subsequently the infinite volume limit $L \to \infty$ has to be carried out. We refer to 
\eqref{gge1} as \textit{generalized Gibbs ensemble}. Infinite volume averages will be denoted by
$\langle \cdot \rangle_\rho$. 

By integrability, the average charge density can be concisely expressed  as \cite{zamolodchikov1990thermodynamic,SALEUR2000602}
 \begin{equation}\label{charge}
     \langle q_i(0)\rangle_\rho=\int_\mathbb{R}\dd\theta h_i(\theta)\rho(\theta).
 \end{equation}
Here $h_i$ characterizes the $i$-th charge, and possible extra indices that label additional particle species are suppressed. For example, for the Lieb-Liniger model, $h_i(\theta)
 = \theta^i$. The function $\rho(\theta)$ is the \textit{root density}, which uniquely characterizes the GGE. This is why we can label each GGE average by $\rho$ in an unambiguous way. Hence it is convenient to
 also use the subscript $\rho$ instead of $\mathrm{GGE}$. 
 There is a somewhat different point of
view. The charge densities form a linear space and \eqref{charge} is a linear functional. Since the set $\{h_i\}$ spans the Hilbert space $L^2(\mathbb{R},\rho(\theta)d\theta)$,
the right hand side represents  this linear functional.


If the system is large and perfectly isolated,
one would expect that the long time behavior is constrained by the conservation laws. To obtain a quantitative theory, we assume 
an initial GGE which has a slow spatial variation as
\begin{equation}\label{inhomo}
Z^{-1}\exp\Big[ - \sum_{i=0}^\infty\int \dd x \beta^i(\epsilon x) q_i(x) \Big].
 \end{equation}
The parameter $\epsilon >0$ is dimensionless and controls the ratio ``microscopic length/\\macroscopic length'', which is assumed to be small. 
The assumption of local equilibrium means that the time-evolved state is again of the form \eqref{inhomo} provided the parameters of the GGE
are properly adjusted in space-time. Such an approximation is valid only  in the sense of equality of expectations of local observables.
With the aim to obtain a closed equation for $\rho$, the conservation law \eqref{conseq} is now averaged over the state at time $t$, $\langle \cdot \rangle_{t,\epsilon}$, with the result
\begin{equation}\label{conseq1}
     \partial_t \langle q_i(x)\rangle_{t,\epsilon}+\partial_x\langle j_i(x)\rangle_{t,\epsilon}=0, \quad i = 0,1,...\,,
 \end{equation}
still an exact identity. Local GGE can be applied, since the operators appearing in \eqref{conseq1} are local. Thus to lowest order in the parameter $\epsilon$,
the average in \eqref{conseq1} refers to a GGE with  chemical potentials depending on the macroscopic space-time point $(\epsilon^{-1}x,\epsilon^{-1}t)$. By this assumption 
the average simplifies dramatically. Required are only the averages
$\langle q_i(0)\rangle_\rho$ and $\langle j_i(0)\rangle_\rho$, where $\rho$ refers to a spatially homogeneous GGE.  The first set of quantities have been addressed already. The GGE averaged currents
will be the main focus of the second part of our review. 
Since eventually one has to come up with a closed set of hydrodynamic equations, the more precise task
is to obtain the GGE averaged currents as a functional of the GGE root density $\rho$.


Current densities also form a linear space. Hence one can  write
\begin{equation}\label{current}
     \langle j_i(0)\rangle_\rho=\int_\mathbb{R}\dd\theta h_i(\theta) v^\mathrm{eff}(\theta)\rho(\theta),
 \end{equation}
for some yet to be determined function $v^\mathrm{eff}(\theta)$, which is a functional of $\rho$. With these notations the equations of GHD
become
\begin{equation}\label{ghdevolution}
     \partial_t\rho(\theta)+\partial_x\big(v^\mathrm{eff}(\theta)\rho(\theta)\big)=0.
 \end{equation}
For simple fluids the Euler currents can be expressed through the pressure as a function of the density and the internal 
 energy, which commonly is referred to as equation of state. Hence the GGE averaged currents in \eqref{ghdevolution} are also dubbed ``equation of state'', although they are no longer directly related to the free energy.
 
 Physically, the expansion in $\epsilon$ corresponds  to a gradient expansion. The Euler equations are of order 
$\epsilon$. Thus, in microscopic units, the spatial variation is order $1/\epsilon$ and one has to observe times also of order
$1/\epsilon$ to detect a motion of the conserved fields. In this sense the Euler scaling is ballistic. The next order in the gradient expansion carries two spatial derivatives and corresponds to the dissipative Navier-Stokes equations,
which have been investigated only recently  \cite{10.21468/SciPostPhys.6.4.049}.\newpage
\noindent
{\Large{\textbf{Part I: Form factors}}}
\section{Form factors and correlation functions in integrable systems}
\subsection{GHD and correlation functions}
Most applications of GHD are focused on the computation of one-point functions. This problem becomes non-trivial when considering spatially inhomogeneous non-equilibrium settings, leading to one-point functions with interesting space-time dependence. This is unlike the more trivial ground-state one-point functions, which by translation invariance are constant. Spatial inhomogeneities make even the calculation of one-point functions a difficult theoretical task. GHD approaches this problem by considering situations in which the spatial inhomogeneity is slowly varying, an approximation which often becomes exact once the inhomogeneous system is left in isolation to evolve over long times. In this slowly varying regime, quantum fluctuations are suppressed, and one-point functions can be computed using hydrodynamic equations. Even though  computing one-point functions for spatial inhomogeneous states is already a worthwhile problem, one may also be interested in obtaining higher-point functions, which are more demanding. In this article the focus will be mostly on two-point functions.

The computation of multi-point functions is a non-trivial problem even for an integrable system in its ground state \cite{doi:10.1142/1115}. The most important and straightforward approach towards computing two-point functions is the spectral decomposition and summation over intermediate matrix elements. Considering a ground-state two-point function of the form
\beq
C^{\mathcal{O}_1\mathcal{O}_2}(x_1,x_2)=\langle 0\vert \mathcal{O}_1(x_1)\mathcal{O}_2(x_2)\vert 0\rangle=\sum_n\langle 0\vert \mathcal{O}_1(x_1)\vert n\rangle\langle n\vert\mathcal{O}_2(x_2)\vert 0\rangle,\label{twopoint}
\eeq
one  inserts a summation over a complete set of eigenstates $\vert n\rangle$ of the Hamiltonian. If  matrix elements $\langle 0\vert\mathcal{O}_{1,2}(x)\vert n\rangle$ can be computed and efficiently summed (or summing only a small number of them for certain approximations), then the two-point function can be calculated. As will be discussed, the problem of computing two-point functions is more complicated for a system  at finite temperatures, or out of equilibrium, for example in a spatially inhomogeneous setting. 

The GHD formalism provides an alternative approach for computing two-point functions, not relying on spectral decomposition \cite{10.21468/SciPostPhys.5.5.054}. This calculation is applicable again only in the regime of slowly varying spatial inhomogeneities. Furthermore it is a long-distance approximation,
in the sense that two operators are well separated in space-time. The GHD computation provides novel useful results even in case of spatially homogeneous 
systems, by providing new predictions for  long-distance two-point functions at finite temperatures. 

The following section is a brief review of traditional matrix-element-based techniques for computing ground-state correlation functions in integrable systems, with a particular focus on quantum field theories. We will discuss the advances with the perspective of generalizing these techniques towards  GGE and non-equilibrium states. Also explained are novel thermodynamic matrix element techniques, which are used to recover the results for two-point functions based on GHD in spatially homogeneous settings, as well as one-point functions in the case of spatial inhomogeneity. Form factors are particularly adapted to integrable systems, since the complete spectrum of eigenstates $\{\vert n\rangle\}$ can be computed exactly using the Bethe Ansatz \cite{korepin_bogoliubov_izergin_1993}.  Furthermore, form factors themselves can be obtained exactly, or at least those involving the lowest lying eigenstates. Both of these aspects are particularly simple for relativistic quantum field theories at infinite volume, so we will primarily focus on this type of model.

\subsection{Form factors in integrable quantum field theories}
In this section we review the standard form factor bootstrap formalism and how it can be applied to compute {\it ground state} correlation functions. This will give an idea of the available tools to study integrable QFT's, as well as provide a necessary background to understand the generalizations to the thermodynamic case we consider in the following sections.

The asymptotic scattering eigenstates in QFT are particle-like, and follow a relativistic dispersion relation. Integrability in QFT requires that all scattering events are  elastic, meaning that the  set of momenta for incoming particles is conserved, and is the same set, up to a re-ordering of particles,  in the outgoing state. From these considerations, one concludes that the eigenstates of an integrable QFT, at infinite volume, are of the form
\beq
\vert \theta_1,\dots,\theta_n\rangle_{\rm in}.
\eeq
Here $\theta_i$ is the rapidity of the $i$-th particle, which parametrizes energy and momentum as $E_i=m_i\cosh\theta_i$, $p_i=m_i\cosh\theta_i$,
according to a relativistic dispersion relation with particle mass $m_i$. For incoming scattering states, $t\to-\infty$, the rapidities are arranged so that $\theta_1>\theta_2>\dots>\theta_n$. In this review, for brevity, we shall deal with diagonally scattering theories with single particle species only, whenever we talk about integrable QFTs.

Quasi-particles states of integrable field theories may also possess internal symmetries, with multiple species of particles. Elasticity requires that, upon scattering, the identity of a particle may change to another species as long as they are of equal mass (this is known as non-diagonal scattering). For the sake of simplicity we focus here on the most elementary kind of integrable field theory, which has only a single species. We then define the S-matrix, $S(\theta)$, as
\beq
\,_{\rm out}\langle \theta_1^\prime,\theta_2^\prime\vert \theta_1,\theta_2\rangle_{\rm in}=(2\pi)^2\delta(\theta_1^\prime-\theta_1)\delta(\theta_2^\prime-\theta_2)S(\theta),
\eeq
where $\theta=\theta_1-\theta_2$. There are further constraints imposed on the S-matrix of an integrable QFT \cite{ZAMOLODCHIKOV1979253}, based on elasticity, factorization, unitarity, relativistic invariance, and internal symmetries of the model. For example, in a model with only one species of  particles, a particle being its own antiparticle, unitarity and crossing symmetry yield the constraints 
\beq
S(\theta)S(-\theta)=1,\,\,\,S(\theta)=S(\pi {\rm i}-\theta),
\eeq
respectively. Such constraints, together with the assumption of minimality (any poles or zeros of the S-matrix for physical values of $\theta$ must correspond to physical bound states or resonances), allow in many cases for a full determination of the S-matrix \cite{mussardo2010statistical}.

The approach towards computing form factors in integrable QFT is comparable. Now strong constraints 
are imposed on form factors. In some cases this yields an exact solution \cite{karowski1978exact,doi:10.1142/1115}. For example, consider the two-particle form factor for a one-particle species QFT, 
\beq
f^\mathcal{O}(\theta_1,\theta_2)(x)=\langle0\vert\mathcal{O}(x)\vert \theta_1,\theta_2\rangle.
\eeq
The operator can be moved to the origin by translating relativistically (assuming $\mathcal{O}$ is a scalar-valued operator, for simplicity) as $f^\mathcal{O}(\theta_1,\theta_2)(x)=e^{{\rm i} x\cdot (p_1+p_2)}f^\mathcal{O}(\theta_1,\theta_2)$. Relativistic invariance implies that the form factor  depends only  on differences of rapidities, $f^\mathcal{O}(\theta_1,\theta_2)=f^\mathcal{O}(\theta)$, where $\theta=\theta_1-\theta_2$. This implies that one-particle form factors, $\langle 0\vert\mathcal{O}\vert \theta_1\rangle$, are constant. We will not provide a complete list of form factor constraints, but only the important ones. 

First in the list is the so-called ``scattering axiom'', which rules the exchange in order of two particles. The axiom states that two particles in the incoming state can be exchanged by multiplying with  the S-matrix as 
\beq
f^\mathcal{O}(\theta)=S(\theta)f^\mathcal{O}(-\theta).\label{scattering}
\eeq
Another property is the so-called ``periodicity axiom'', based on crossing symmetry, which states that if one crosses particles to the outgoing state and then crosses them again to the incoming state, the form factor is invariant. The crossing operation consists ot transforming an incoming particle into an outgoing anti-particle, which kinematically amounts to reversing the sign of energy and momenta of that particle in the S-matrices or form factors. In terms of rapidities, crossing is achieved by the transformation $\theta\to\theta+\pi{\rm i}$. When applying a crossing transformation on the second particle, to become an  anti-particle in the outgoing state, and then applying another crossing transformation, bringing the particle back to the incoming state, it results in the consistency condition,
\beq
f^\mathcal{O}(\theta)=\gamma^{\mathcal{O}_2}f^\mathcal{O}(2\pi {\rm i}-\theta).\label{periodicity}
\eeq
Here $\gamma^{\mathcal{O}_2}$ is a ``semi-locality'' proportionality factor between the operator and the second particle being crossed around the operator. This factor is 1 only for strictly local excitations, but can be different for semi-local excitations, such as topological solitons.  Given an S-matrix, one can solve equations (\ref{scattering},\ref{periodicity})  exactly in the form 
\beq
f^\mathcal{O}(\theta)=g^\mathcal{O}(\theta)f^{\rm min}(\theta).\label{minimalform}
\eeq
The factor $f^{\rm min}(\theta)$ is the minimal solution of equations (\ref{scattering},\ref{periodicity}), such that there are no physical poles or zeros.
$f^{\rm min}(\theta)$ is operator-independent. $g^\mathcal{O}$ is a periodic operator-dependent factor, $g^\mathcal{O}(\theta)=g^\mathcal{O}(\theta+2\pi {\rm i})$, which contains all required singularities.

Equation (\ref{minimalform}) is easy to solve for general QFT's with one species of particle. If we express the two-particle S-matrix in terms of an auxiliary function $B(t)$, defined by
\beq
S(\theta)=\exp\left[\frac{1}{2}\int_{-\infty}^\infty\frac{dt}{t}B(t)\exp\frac{t\theta}{{\rm i}\pi}\right],
\eeq
then the minimal form factor can be found to equal (\cite{karowski1978exact}),
\beq
f^{\rm min}(\theta)=\exp\left[-\frac{1}{4}\int_{-\infty}^\infty\frac{dt}{t}\frac{B(t)}{\sinh t}\exp\left(\frac{t(\theta-{\rm i}\pi)}{{\rm i}\pi}\right)\right].
\eeq

A particular example of such a one-species QFT is the sinh-Gordon model, with bosonic field $\phi$ and coupling $g$, defined by the action,
\beq
\mathcal{S}=\int d^2x\left(\frac{1}{2}(\partial_\mu\phi(x))^2-\frac{m^2}{g^2}\cosh g\phi(x)\right).
\eeq
The 2-particle S matrix is \cite{arinshtein1979quantum},
\beq
S(\theta)=\frac{\tanh\frac{1}{2}(\theta-{\rm i}\pi b/2)}{\tanh \frac{1}{2}(\theta+{\rm i}\pi b/2)},
\eeq
with $S(0)=-1$, and 
\beq
b=\frac{2g^2}{8\pi+g^2}.
\eeq
An often studied local observable is the  vertex operator, $\mathcal{O}_\alpha\equiv e^{{\rm i}\alpha\phi}$. For these operators, it can be shown the 2-particle form factors are given simply by  the minimal form factor, times a $\alpha$-dependent normalization constant \cite{lukyanov1997form}, $f_\alpha(\theta)=K_\alpha f^{\rm min}(\theta)$.

We will only briefly touch the role of singularities in form factors. As mentioned, poles in the S-matrix (and form factors) may correspond to the existence of particle bound states. There also exist annihilation poles, where a particle and antiparticle in the crossed channel may annihilate when their rapidities coincide. For a general $n$-particle form factor, this implies
\beq
-{\rm i}{\rm Res}_{\theta_{n-1}+\pi {\rm i}=\theta_n}f^\mathcal{O}(\theta_1,\dots,\theta_n)=\left(1-\gamma^{\mathcal{O}n}S(\theta_n-\theta_1)\dots S(\theta_{n}-\theta_{n-2})\right) f^\mathcal{O}(\theta_1,\dots,\theta_{n-2}).
\eeq
This form factor provides a recursive relation between the $n$ and $n-2$ particle form factors. The case of a two-particle form factor is particularly sensitive to the semi-locality factor, since the residue vanishes and there is no pole when $\gamma^{\mathcal{O}_2}=1$. The two-particle form factor has only an annihilation pole when the particles are semi-local with respect to the operator \cite{delfino1996non}. 

Based on these, and other form factor axioms, thoroughly discussed in \cite{doi:10.1142/1115,mussardo2010statistical}, one readily computes the few-particle form factors of many QFT's and therefore also (approximate) two-point correlation functions in the ground state.

\subsection{A non relativistic example: The Lieb-Liniger model}

The form factor bootstrap formalism as described above is a powerful tool for obtaining (or approximating) correlation functions in the ground state of an 
integrable quantum field theory. Several of the form factor axioms rely on relativistic invariance, which means that they are not applicable to 
non-relativistic integrable continuum models or quantum spin chains. However, in those cases one may have knowledge of form factors between the ground and  excited states by other means \cite{korepin_bogoliubov_izergin_1993}. 

A prototypical well-studied  example of a non-relativistic integrable system is the Lieb-Liniger model, with Hamiltonian,
\beq
H=-\sum_{j=1}^N\partial_{x_j}^2+2c\sum_{1 \leq j<k \leq N}\delta(x_j-x_k).
\eeq 
We will focus on the repulsive case, $c>0$. From our QFT point of view, this model arises from taking the non-relativistic limit of the sinh-Gordon field theory, with a fixed number of particles \cite{Kormos_2010}. There is in fact recent evidence that the non-relativistic limit of {\it any} integrable QFT is described by a generalized version of the Lieb-Liniger model \cite{bastianello2016non}.

Lieb-Liniger form  factors can  be computed through Algebraic Bethe ansatz methods \cite{slavnov1990nonequal}. Fixing the particle number to $N$, and placing the system in a finite volume, $L$, with periodic boundary conditions, unnormalized eigenstates are specified as $\vert \{\lambda\}\rangle$, where $\{\lambda\}$ is a set of $N$ particle rapidities, determined by a solution of the Bethe  equations. In this model the rapidities parametrize energy and momentum as $p(\lambda)=\lambda$, $E(\lambda)=\lambda^2$.  The norm of such states is given by \cite{gaudin1971bose,korepin1982calculation}
 \beq
 \langle \{\lambda\}\vert\{\lambda\} \rangle=c^N\prod_{j\neq k}^N\frac{\lambda_j-\lambda_k+{\rm i}c}{\lambda_j-\lambda_k}{\rm det}_N G,
 \eeq
 with the Gaudin matrix
 \beq\label{gaudinll}
 G_{jk}&=&\delta_{jk}\left(L+\sum_{m=1}^N\varphi_\mathrm{LL}(\lambda_j-\lambda_m)\right)-\varphi_\mathrm{LL}(\lambda_j-\lambda_k)
 \eeq
 and the Lieb-Liniger scattering shift
 \beq
 \varphi_\mathrm{LL}(\lambda)&=&\frac{2c}{\lambda^2+c^2}.
 \eeq
 One example of known form factors concerns the particle density operator, defined as $\hat{\rho}(x)=\sum_{j=1}^N\delta(x-x_j)$. In this case, the matrix element between two arbitrary Bethe states of the same $N$ is given by \cite{slavnov1990nonequal}
 \beq
 \langle\{\mu\}\vert\hat\rho(0)\vert\{\lambda\}\rangle=\frac{\partial}{\partial\alpha}\sigma_N^\alpha(\{\mu\},\{\lambda\})_{\alpha=0},\label{generalformll}
 \eeq
 where
 \beq
 \sigma_N^\alpha(\{\mu\},\{\lambda\})=\frac{\prod_{j,m}(\lambda_{jm}+{\rm i}c)}{\prod_{j>m}\lambda_{jm}\mu_{mj}}\,{\rm det}_{N}\left(M_{jk}^{(1)}+M_{jk}^{(2)}\right),
 \eeq
 \beq
 M_{jk}^{(1)}=\frac{V_j^+e^\alpha-V_j^-}{\mu_k-\lambda_j},\,\,\,\,M_{jk}^{(2)}=\frac{V_j^-}{\mu_k-\lambda_j-{\rm i}c}-\frac{e^\alpha V_j^+}{\mu_k-\lambda_j+{\rm i}c},
 \eeq
 and
 \beq
 V_j^{\pm}=\prod_{m=1}^N\frac{\mu_m-\lambda_j\pm{\rm i}c}{\lambda_m-\lambda_j\pm{\rm i}c},
 \eeq
 where we used the notation $\lambda_{jk}=\lambda_j-\lambda_k$.
 
 Note that these form factors differ from the ones of QFT, since the particle number is necessarily the same for both incoming and outgoing states due to the non-relativistic nature of the model (particles cannot be spontaneously created or destroyed).

\subsection{Quantum spin chains}

Form factor methods can also be applied to integrable quantum spin chains. For simplicity,  our discussion will be restricted to the transverse field Ising chain (TFIC), whose Hamiltonian is given by
\beq
H=-J\sum_{j=1}^L(\sigma_j^x\sigma_{j+1}^x+h\sigma_j^z).
\eeq
This model is well known to be solvable via Jordan-Wigner and Bogoliubov transformation \cite{schultz1964two}. The eigenstates are described by a set of quasiparticles with momentum $k$ and energy $\varepsilon(k)=2J\sqrt{1+h^2-2h\cos k}$. Furthermore, the Hilbert space is split into a 
Neveu-Schwartz (NS) and a Ramond (R) sector. In the NS (R) state, values of momenta are quantized, having half-integer (integer) multiples of $\frac{2\pi}{L}$. Form factors can be computed for operators such as the Pauli matrices $\sigma^x_j,\sigma^z_j$. The TFIC quasi-particles are free and fermionic, in the sense that their scattering matrix is -1. This renders the form factors of the operator $\sigma^z_j$ trivial, being non-zero only for two-particle form factors. 

Due to the Jordan-Wigner transformation, the particle excitations are non-local relative to the operator $\sigma^x_j$ (the particles have a semi-locality factor $\gamma^{\sigma^x n}\neq 1$). This generates non-trivial form factors, as well as annihilation poles at the two-particle form factor level.  The  $\sigma^x_j$ operator has non-vanishing form factors only between states in opposite sectors (NS and R).  For $\sigma^x_j$ the general form factor can be written as 
\beq
&&\hspace{-26pt}\,_{\rm NS}\langle q_1,\dots,q_{2n}\vert \sigma_0^x\vert p_1,\dots,p_m\rangle_R={\rm i}^{[n+m/2]}(4J^2h)^{(m-2n)^2/4}\sqrt{\xi\xi_L}\nonumber\\
&&\hspace{-16pt}\times\prod_{j=1}^{2n}\left(\frac{e^{\eta_{q_j}}}{L\varepsilon(q_j)}\right)^{1/2}\prod_{l=1}^m\left(\frac{e^{\eta_{p_l}}}{L\varepsilon(p_l)}\right)^{1/2}\prod_{j<j^\prime}^{2n}\frac{\sin\frac{q_j-q_{j^\prime}}{2}}{\varepsilon_{q_jq_{j^\prime}}} \prod_{l<l^\prime}^{m}\frac{\sin\frac{p_l-p_{l^\prime}}{2}}{\varepsilon_{p_lp_{l^\prime}}}\prod_{j=1}^{2n}\prod_{l=1}^m\frac{\varepsilon_{q_jp_l}}{\sin\frac{q_j-p_l}{2}},\label{isingformgeneral}
\eeq
where $\xi=\vert 1-h^2\vert^{1/4}$, $m$ is even (odd) for $h<1$ $(h>1)$, and $\varepsilon_{ab}=(\varepsilon(a)+\varepsilon(b))/2$ \cite{bugrii2001correlation,bugrij2003spin,von2008form,iorgov2011spin}.  For large $L$ one can use the approximations $\xi_L\sim 1$, $e^{\eta_k}\sim 1$. Note that the two-particle form factor (for $n=0$, and $m=2$, with $h<1$) has an annihilation pole, due to the semi-locality of kink excitations.

\section{Thermodynamics of integrable systems}\label{tba}

\subsection{Thermodynamic Bethe ansatz}
Thermodynamics of integrable systems is known to be  described by the thermodynamic Bethe ansatz (TBA)\cite{Yang1969,zamolodchikov1990thermodynamic}. The structure of TBA is universal and essentially independent of the particular quantum integrable system under consideration. For simplicity, we show the details of the TBA formalism focusing on QFT's with one species of particle and  two-body scattering matrix $S(\theta)$. 

An important use of the TBA is to evaluate the free energy $F=-\log Z$ in a generalized Gibbs ensemble, where we recall that the partition function $Z$ is given by
\begin{equation}
    Z=\mathrm{Tr}\big(\exp\big[ -\sum_{i=0}^\infty\beta^iQ_i\big]\big),
\end{equation}
compare with \eqref{gge1}.
 In our context ``thermodynamic'' should be really understood as ``generalized thermodynamics'', in the sense that higher labeled  conserved charges are controlled by chemical potentials, whereas traditionally a thermodynamic partition function has only inverse temperature and the chemical potential for particle number
 as control parameters. Nevertheless we use the term ``thermodynamic'' henceforth.

Assume that the system is put on a ring with circumference $L$. When there are $n$ quasi-particles, the momenta are quantized according to the Bethe equation $e^{-\ii Lp(\theta_i)}=\prod_{j\neq i}^nS(\theta_i-\theta_j)$, which immediately gives the Bethe-Yang equations
\begin{equation}\label{betheyang}
    \psi_i(\theta_1,\dots,\theta_n)=p(\theta_i)L+\sum_{j=1, j\neq i}^n\phi(\theta_i-\theta_j)=2\pi  I_i,
\end{equation}
where $\phi(\theta)=-\ii\log S(\theta)$ is the phase shift and $I_i\in\mathbb{Z}$. In the infinite volume limit $L\to\infty$, the distribution of quasi-particles are controlled by the root density $\rho(\theta_j)=\lim_{L\to\infty}(L(\theta_{j+1}-\theta_j))^{-1}$, where $\theta_j<\theta_{j+1}$ are two adjacent Bethe roots, i.e. solutions of the Bethe equation. The equation satisfied by the root density can be inferred by taking the derivative of the thermodynamic limit of the Bethe-Yang equations, with the result
 \begin{equation}\label{statedensity}
     \rho^\tot(\theta)=\frac{p'(\theta)}{2\pi}+\int_\mathbb{R}\frac{\dd\theta'}{2\pi}\varphi(\theta-\theta')\rho(\theta'),
 \end{equation}
where $\rho^\tot(\theta)$ is the state density and $\varphi(\theta)= \phi'(\theta)$ is the scattering phase shift.  Since excitations can be either particle or hole, the state density is given by $\rho^\tot(\theta)=\rho(\theta)+\rho^\mathrm{h}(\theta)$, where $\rho^\mathrm{h}(\theta)$ designates the hole density. At finite temperatures, i.e. for some arbitrary GGE, one can fix the relation between $\rho(\theta)$ and $\rho^\mathrm{h}(\theta)$ by requiring the minimization of the free energy $F[\rho]=\sum_i\beta^iQ_i[\rho]-TS[\rho]$. Thereby it follows that the occupation function, defined as $n(\theta)=\rho(\theta)/\rho^\tot(\theta)$, satisfies 
\begin{equation}\label{gge}
    \varepsilon(\theta)=\sum_{i= 0}^\infty\beta^ih_i(\theta)-\int_\mathbb{R}\frac{\dd\theta'}{2\pi}\varphi(\theta-\theta')\log(1+e^{-\varepsilon(\theta')}).
\end{equation}
Here $h_i(\theta)$ is the one-particle eigenvalue of $Q_i$ and the pseudo-energy, $\varepsilon(\theta)$, is defined by
\beq
\frac{\rho(\theta)}{\rho^{\rm tot}(\theta)}=\frac{e^{-\varepsilon(\theta)}}{1+e^{-\varepsilon(\theta)}}.
\eeq
In terms of the pseudo-energy, the free energy can be expressed as
\begin{equation}
    F=-L\int_\mathbb{R}\frac{\dd\theta}{2\pi}p'(\theta)\log\left(1+e^{-\varepsilon(\theta)}\right).
\end{equation}

For our purposes it is important to observe that the GGE average of charge densities $q_i$ 
can be obtained by taking the derivative of the free energy with respect to $\beta^i$, i.e. $\p F/\p\beta^i=\langle Q_i\rangle=L\langle q_i\rangle$, from which one infers 
\begin{equation}\label{chargeaverage}
     \langle q_i(0)\rangle_\rho=\int_\mathbb{R}\dd\theta h_i(\theta)\rho(\theta),
 \end{equation}
in agreement with \eqref{charge}.

At finite volume, according to (\ref{betheyang}), when a new particle (or hole) is added to the system, the rapidities of all other particles are modified and this change is of order $1/L$. Such an effect can be quantified in the infinite volume limit, in terms of the ``backflow'' function,  $\mathcal{F}(\tilde\theta\vert\theta)$, which is defined by the change in the rapidity $\tilde\theta$ of a background particle, when a new particle of rapidity $\theta$ is added,
\beq
\tilde\theta\to\tilde\theta-\frac{\mathcal{F}(\tilde\theta\vert\theta)}{L\rho^\tot(\theta)}.
\eeq
In the infinite volume limit, this backflow function can be shown \cite{korepin_bogoliubov_izergin_1993} to be given by the solution of
\beq
2\pi\mathcal{F}(\tilde\theta\vert\theta)=\phi(\tilde\theta-\theta)+\int_\mathbb{R} \dd\theta^\prime \varphi(\tilde\theta-\theta^\prime)n(\theta')\mathcal{F}(\theta^\prime\vert\theta).\label{backflow}
\eeq
While the shift of each background particle rapidity is of order $1/L$, there are $N\sim L$ background particles, so the total shift of particle rapidities produces a finite effect.  Using the backflow function, one defines quantities such as ``effective'' energy and momentum of the added particles. For example, in relativistic QFT, if a particle with rapidity $\theta$ is added, then its effective energy and momentum resulting from the total shift of the background is given by
\beq
p_{\rm eff}(\theta)&=&p(\theta)-m\int_\mathbb{R}\dd\theta^\prime\cosh\theta^\prime n(\theta^\prime)\mathcal{F}(\theta^\prime\vert\theta),\nonumber\\
E_{\rm eff}(\theta)&=&E(\theta)-m\int_\mathbb{R} \dd\theta^\prime\sinh\theta^\prime n(\theta^\prime)\mathcal{F}(\theta^\prime\vert\theta).\label{effectiveep}
\eeq
We remark that the $\theta$ derivatives of $p_{\rm eff}(\theta)$ and $E_{\rm eff}(\theta)$ coincide with the dressed versions of $p'(\theta)$ and $E'(\theta)$, i.e. $\p_\theta p_{\rm eff}(\theta)=(p')^\dr(\theta)$ and $\p_\theta E_{\rm eff}(\theta)=(E')^\dr(\theta)$ where the dressing operation is defined for an arbitrary function $h(\theta)$ as
\begin{equation}\label{dressing}
    h^\dr(\theta)=h(\theta)+\int_\mathbb{R}\frac{\dd\theta'}{2\pi}\varphi(\theta,\theta')n(\theta')h^\dr(\theta').
\end{equation}
Here we assumed that $\varphi(\theta,\theta')$ is symmetry with respect to an exchange of $\theta$ and $\theta'$.

\subsection{LeClair-Mussardo formula and thermodynamic representative states}


A thermodynamic state is characterized by a rapidity distribution, $\rho(\theta)$, which is used as an input to the TBA equations to yield the free energy. The computation of expectation values of local operators generally requires the knowledge of form factors in addition to TBA equations. For instance general one-point functions in QFT can be computed at finite temperature using the Leclair-Mussardo formula \cite{LMformula}, which uses as an input the standard ground-state form factors from the previous section. 


For now we focus on the case of a generic GGE. The expectation values of local operators are given by
\beq\label{gge2}
\langle\mathcal{O}\rangle_\mathrm{GGE}=\mathrm{Tr}(\varrho\mathcal{O}),
\eeq
where the normalized GGE density matrix, $\varrho$, has been defined in \eqref{gge1}.
The difficult trace can be avoided by choosing a \textit{representative eigenstate} 
of the Hamiltonian with the property that its expectation values of local observables agree with \eqref{gge} in the infinite volume limit \cite{korepin_bogoliubov_izergin_1993}. 

We  provide here a definition of such representative states for  QFT, though they can be similarly defined for non-relativistic systems and spin chains. One starts from QFT eigenstates in a finite volume, $L$. An $N$-particle state is characterized by the set of $N$ rapidities, $\{\tilde{\theta}\}$ as a solution of the Bethe-Yang equations \eqref{betheyang} for a specified set of $N$ integers $\{I\}$. To emphazise the fact that eigenstates are specified by a set of integers, instead of continuous rapidities, one can switch to an integer eigenstate basis defined by
\beq
\vert I_1,\dots,I_n\rangle=\frac{1}{\sqrt{\det G(\tilde{\theta}_1,\dots,\tilde{\theta}_n)}}\vert\tilde{\theta}_1,\dots,\tilde{\theta}_n\rangle,
\eeq
where the Gaudin matrix $G_{jk}(\tilde{\theta}_1,\dots,\tilde{\theta}_n)$
is the Jacobian of the transformation between integers and rapidities, as obtained from the Bethe-Yang equations,
\begin{equation}\label{gaudin}
    G_{jk}(\theta_1,\dots,\theta_n)=\frac{\p\psi_k(\theta_1,\dots,\theta_n)}{\p\theta_j}=\Big((Lp'(\theta_j)+\sum_{l=1}^n\varphi(\theta_j-\theta_l)\Big)\delta_{jk}-\varphi(\theta_k-\theta_j).
\end{equation}
Note that this expression coincides with the Gaudin matrix in \eqref{gaudinll} by substituting the scattering shift as  $\varphi=\varphi_\mathrm{LL}$. A representative state is defined by taking the thermodynamic limit of a finite volume eigenstate, such that the set of rapidities $\{\theta\}$ are distributed according to the root density $\rho(\theta)$ of the GGE in \eqref{gge}. We use $|\rho\rangle$ to denote such a unnormalized representative state. Then, in approximation, $\langle\mathcal{O}\rangle_\mathrm{GGE}=\langle\rho|\mathcal{O}\vert\rho\rangle / \langle\rho\vert\rho\rangle$ for local observables.
The representative state is not unique, as one can make non-extensive changes to the set of rapidities (like adding or removing a finite number of particles), which gives a state $\vert \{I^\prime\}\rangle$ that still yields in the infinite volume  limit the same distribution. The number of such states that are described by the same distribution is given in the thermodynamic limit by, $e^{S_{YY}[\rho]}$, where $S_{YY}[\rho]$ is the Yang-Yang entropy \cite{Yang1969}.



Given a representative state $\vert \rho\rangle$,  
%
%
 it can be shown that expectation values of local operators can be evaluated exactly as an expansion in terms of the standard ground-state form factors. This formula was first proposed by Leclair and Mussardo \cite{LMformula}, later proven to be correct in the case for conserved charges in \cite{SALEUR2000602} and for generic cases in \cite{Pozsgay_2011,10.1007/978-981-13-2179-5_6}.  The Leclair-Mussardo formula states
\beq\label{LM}
\langle\mathcal{O}\rangle_\rho=\frac{\langle \rho\vert \mathcal{O}\vert\rho\rangle}{\langle \rho\vert\rho\rangle}=\sum_{n=0}^\infty\frac{1}{n!}\int_{\mathbb{R}^n}\prod_{i=1}^n\left[\frac{\dd\theta_i}{2\pi}n(\theta_i)\right]f^{\mathcal{O}}_\mathrm{c}(\theta_1,\dots,\theta_n),\label{leclairmussardo}
\eeq
where the connected form factor is defined as
\beq
f^\mathcal{O}_\mathrm{c}(\theta_1,\dots,\theta_n)=\lim_{\{\xi_i\}\to0}\langle0\vert\mathcal{O}\vert\theta_1,\dots,\theta_n,\theta_n-{\rm i}\pi+{\rm i}\xi_n,\dots,\theta_1-{\rm i}\pi+{\rm i}\xi_1\rangle\vert_{\rm finite\,part}
\eeq
and by ``finite part'' means to remove all divergent terms when taking the limit $\{\xi_i\}\to0$.

Alternatively one can express the LeClair-Mussardo formula \eqref{LM} as \cite{Pozsgay_2011}
\beq\label{LM2}
\langle\mathcal{O}\rangle_\rho=\sum_{n=0}^\infty\frac{1}{n!}\int_{\mathbb{R}^n}\prod_{i=1}^n\left[\frac{\dd\theta_i}{2\pi}n(\theta_i)w(\theta_i)\right]f^{\mathcal{O}}_\mathrm{s}(\theta_1,\dots,\theta_n),
\eeq
where, following \cite{POZSGAY2008209},  we introduced the symmetric form factor, which relies on the alternative regularization
\beq
f^\mathcal{O}_\mathrm{s}(\theta_1,\dots,\theta_n)=\lim_{\{\xi_i\}=\xi\to0}\langle0\vert\mathcal{O}\vert\theta_1,\dots,\theta_n,\theta_n-{\rm i}\pi+{\rm i}\xi,\dots,\theta_1-{\rm i}\pi+{\rm i}\xi\rangle,
\eeq
with weight
\begin{equation}
    w(\theta)=\exp\left(-\int \dd\theta'n(\theta')\varphi(\theta-\theta')\right).
\end{equation}
As proven in \cite{POZSGAY2008209}, in terms of these form factors, the finite-volume diagonal matrix element can be written as
\begin{align}
    \label{diag}
{}_L\langle\tilde{\theta}_n,\dots,\tilde{\theta}_1|\mathcal{O}(0)|\tilde{\theta}_1,\dots,\tilde{\theta}_n\rangle_L&=\sum_{\substack{\alpha \subset \{1,\dots,n\}\\
\alpha\neq\varnothing}}f^\mathcal{O}_\mathrm{s}(\{\tilde{\theta}_i\}_{i\in\alpha})\det G(\{\tilde{\theta}_i\}_{i\in\bar{\alpha}})+\mathcal{O}(e^{-\mu L})\n
&=\sum_{\substack{\alpha \subset \{1,\dots,n\}\\
\alpha\neq\varnothing}}f^\mathcal{O}_\mathrm{c}(\{\tilde{\theta}_i\}_{i\in\alpha}) G(\alpha|\alpha)+\mathcal{O}(e^{-\mu L}).
\end{align}
Here $\mu$ is a mass scale, which depends on microscopic details of the theory, and $\bar{\alpha}$ is the complement of the set $\alpha$. $G(\alpha|\alpha)$ is the principal minor of the Gaudin matrix \eqref{gaudin} obtained by removing rows and columns corresponding to the integer set $\alpha$ in the matrix. In fact, the following relation turns out to be algebraic, i.e. holds for any value of $L$,
\begin{equation}\label{diag2}
  \sum_{\substack{\alpha \subset \{1,\dots,n\}\\
\alpha\neq\varnothing}}f^\mathcal{O}_\mathrm{s}(\{\theta_i\}_{i\in\alpha})\det G(\{\theta_i\}_{i\in\bar{\alpha}})=\sum_{\substack{\alpha \subset \{1,\dots,n\}\\
\alpha\neq\varnothing}}f^\mathcal{O}_\mathrm{c}(\{\theta_i\}_{i\in\alpha})G(\alpha|\alpha).
\end{equation}
This identity allows one to express a symmetric form factor $f^\mathcal{O}_\mathrm{s}(\theta_1,\dots,\theta_n)$ just in terms of connected form factors $f^\mathcal{O}_\mathrm{c}(\theta_1,\dots,\theta_n)$ by taking the limit $L\to 0$ \cite{10.21468/SciPostPhys.6.2.023}, 
\begin{equation}\label{symmconn}
f^\mathcal{O}_\mathrm{s}(\theta_1,\dots,\theta_n)=\sum_{\substack{\alpha \subset \{1,\dots,n\}\\
\alpha\neq\varnothing}}L(\alpha|\alpha)f^\mathcal{O}_\mathrm{c}(\{\theta_i\}_{i\in\alpha}),
\end{equation}
 where $L(\alpha|\alpha)$ is the principal minor of the Laplace matrix obtained in the same way as in the Gaudin matrix
\begin{equation}\label{laplace}
    L_{jk}(\theta_1,\dots,\theta_n)=\delta_{jk}\sum_{l = 1,l\neq j}^n\varphi(\theta_j-\theta_l)-(1-\delta_{jk})\varphi(\theta_k-\theta_j).
\end{equation}
This transformation rule turns to be instrumental in one of the proofs of collision rate ansatz later. We emphasize that we need both the connected and symmetric form factors in the proof.
In general it is out of reach to obtain the exact symmetric and connected form factors for a given local operator. There are however some cases for which such a task has been accomplished. For example, the connected form factor of the density of a conserved charge $Q_i$ has been shown to be given by 
\begin{equation}\label{chargeconnff}
    f^{q_i}_\mathrm{c}(\theta_1,\dots,\theta_n)=h_i(\theta_1)\varphi_{1,2}\varphi_{2,3}\cdots\varphi_{n-1,n}p'(\theta_n) + \mathrm{perm.},
\end{equation}
where $\varphi_{i-1,i}=\varphi(\theta_{i-1}-\theta_i)$ and the permutations are with respect to the set $\{1,\dots,n\}$ of integers. It is readily checked that the LeClair-Mussardo formula upon substituting \eqref{chargeconnff} reproduces the TBA expression of the GGE  averaged charge  \eqref{chargeaverage}.


\subsection{Form factor approach to thermodynamic correlation functions}
Having introduced the notions of form factors and ground-state correlation functions, this section combines these ideas, with the aim of computing thermodynamic correlation functions through an analogous expansion in terms of \textit{thermodynamic form factors}. The approach for computing correlation functions through the representative state formula is similar to the ground state case. We simply sum over intermediate states between each operator,
\beq
\langle \rho\vert \mathcal{O}(x_1)\mathcal{O}(x_2)\vert \rho\rangle=\sum_n\langle \rho\vert \mathcal{O}(x_1)\vert n\rangle\langle n\vert\mathcal{O}(x_2)\vert \rho\rangle.
\eeq
The challenge is then to compute form factors of operators between representative states (instead of  ground states as in the previous section) and generic excited states.

The generic form factor $\langle \rho\vert \mathcal{O}(x)\vert n\rangle$ can be simplified through some physical considerations. When the operator $\mathcal{O}(x)$ is local, its action on a representative state $\vert \rho\rangle$ changes this state only locally. This means that the state $\mathcal{O}(x)\vert \rho\rangle$ has a non-vanishing overlap  only with states described by the same distribution $\rho(\theta)$, in other words by  a different realization of the same representative state.  We can then restrict ourselves to states which are excitations of the representative state $\vert \rho\rangle$, rather than excitations relative to the ground states. The representative state 
 can be perturbed either adding or removing particles, or simply changing the momentum of one of the particles in the state (adding a particle-hole pair). The relevant form factors are of the form $\langle \rho\vert \mathcal{O}(x)\vert \bar{\rho}; \theta_1,\dots,\theta_n\rangle$, where $
 \bar{\rho}$ denotes the shifted representative state with shifts in the background particles induced by the additional particles and holes. The energy and momentum carried by these particle/hole excitations are thus dressed by the representative state. This can be understood through the Bethe equations \eqref{betheyang}, since adding a single particle or hole to a finite volume state affects the rapidities of all other particles in the background.  For instance, the effective energy and momenta are modified as per (\ref{effectiveep}). This concept of dressing particle's properties by the thermodynamic background is crucial to our recent understanding of GGE correlation functions. Using this interpretation, one can think of $\langle \rho\vert \mathcal{O}(x)\vert \bar{\rho}; \theta_1,\dots,\theta_n\rangle$ as a dressed version of the ground state form factor $\langle 0\vert \mathcal{O}(x)\vert \theta_1,\dots,\theta_n\rangle$.

\section{Thermodynamic form factors and correlation functions}
\subsection{Quantum field theory}
In the case of standard ground-state form factors, integrability combined with relativistic invariance highly constrains QFT form factors. Typically one is able to compute form factors for a small number of particles by solving the set of form factor axioms. In this section the recent progress obtained in \cite{cubero2019thermodynamic} is discussed, where such form factor axioms are generalized to the case of a thermodynamic background. A similar approach for the computation of thermodynamic form factors in the Ising field theory (a non-interacting fermionic QFT) was previously developed in \cite{Doyon_2007}, yet such methods have not been generalized to interacting QFT's. 

The main contribution in \cite{cubero2019thermodynamic} is to rederive the form factor axioms for particles on top of a thermodynamic background. While ground-state form factors have only particle excitations, for thermodynamic form factors  particles can also be removed from the background, thereby generating holes. In relativistic QFT,  introducing a hole with rapidity $\theta$ is  energetically equivalent to introducing a particle with rapidity $\theta+{\rm i}\pi$. Here we merely illustrate how the axioms are generalized, while a more detailed discussion can be found in \cite{cubero2019thermodynamic}. For this purpose, the simplest possible case to be considered is  a one-particle form factor
\beq
f^\mathcal{O}_\rho(\theta)=\frac{\langle\rho\vert \mathcal{O}\vert \bar{\rho},\theta\rangle}{\langle\rho\vert\rho\rangle}.
\eeq
We point out that a different definition of these form factors has been used in the literature, particularly in \cite{de2018particle}, which includes an "entropic factor", $\exp\left(\frac{1}{2}\delta S[\rho,\{\theta\}]\right)$, which accounts for the fact that there are many microscopic configurations of particle rapidities that lead to the same macroscopic form factor. This entropic factor is given by the change in entropy of the state $\vert\rho\rangle$ caused by adding a set of particles of rapidities $\{\theta\}$. This counts the number of microscopic states that are macroscopically equivalent. We will not include this factor in the rest of our discussion, specially as it doesn't affect our GHD related discussion, since we will mainly be interested in form factors with  particle-hole pair with equal rapidities, in which case the facor becomes one.

Note that there is already an important difference between this thermodynamic form factor and its ground state analog. Relativistic invariance no longer ensures this quantity to be constant. Such form factors are not necessarily invariant under a boost $\theta\to\theta+\alpha$, since the background particles provide a fixed frame of reference.

Also the periodicity axiom is modified by the thermodynamic background. As the particle with rapidity $\theta$ is crossed to the outgoing and then back to the incoming state, it scatters with particles in the thermodynamic background, both in incoming and outgoing state. This product of scattering matrices would nearly cancel each other, if it were not for background particles in the incoming state being slightly shifted by the excitation. As the particle is crossed around the operator, it acquires a phase depending on this shift (for simplicity, we are now concentrating  on strictly local operators with trivial semilocality factor)
as
\beq \label{dressedcrossing}
f^\mathcal{O}_\rho(\theta)=R_\rho(\theta\vert\theta)f_\rho^\mathcal{O}(\theta+2\pi i).
\eeq
 Here
\beq
R_\rho(\theta_i\vert\theta_j)=\exp\big[{\rm i}(2\pi \mathcal{F}(\theta_i\vert \theta_j)-\phi(\theta_i-\theta_j))\big]
\eeq
and $\mathcal{F}(\theta_i\vert\theta_j)$ is the backflow function defined by (\ref{backflow}). The expression (\ref{dressedcrossing}) can be directly contrasted with the ground state version (\ref{periodicity}). Eq. (\ref{dressedcrossing}) already provides a strong restriction, which can be solved in terms of a minimal form factor and an operator-dependent factor analogously to Eq. (\ref{minimalform}).

Explicitly, we can re-express $R_\rho(\theta\vert\theta)$ in exponential form as
\beq
R_\rho(\theta\vert\theta)=\exp\left[\frac{1}{2}\int_{-\infty}^\infty\frac{dt}{t}C_\rho(t)\exp\frac{t\theta}{{\rm i}\pi}\right],
\eeq
such that the general solution of Eq. (\ref{dressedcrossing}) is
\beq
f_\rho^\mathcal{O}(\theta)=K_\rho^\mathcal{O}(\theta)f^{\rm min}(\theta),
\eeq
where 
\beq
f^{\rm min}=\exp\left[-\frac{1}{4}\int_{-\infty}^\infty\frac{dt}{t}\frac{C_\rho(t)}{\sinh t}\exp\left(\frac{t(\theta-{\rm i}\pi)}{{\rm i}\pi}\right)\right]
\eeq
and $K_\rho^\mathcal{O}(\theta)$ is a function periodic under $\theta\to\theta+2\pi {\rm i}$.

In the concrete example of the sinh-Gordon model, it can be shown \cite{cubero2019thermodynamic} that for a vertex operator the quantity $K_\rho^{\mathcal{O}_\alpha}(\theta)$ is just a $\theta$-independent constant, given by 
\beq
K_\rho^{\mathcal{O}_\alpha}=\frac{1}{\langle 0\vert \mathcal{O}_\alpha\vert 0\rangle}\frac{\langle\rho\vert\mathcal{O}_\alpha\vert\vartheta\rangle}{\langle\rho\vert\rho\rangle}f_1^{\mathcal{O}_\alpha},
\eeq
which involves the ground state expectation value, the one-point function in the thermodynamic background, and the ground state one-particle form factor (which is a constant) of the operator.

Unlike the case of ground state correlation functions, expanding the two-point function up to only one-particle form factors already yields interesting dynamical information, since one-particle form factors are not constant. As discussed in \cite{cubero2019thermodynamic} this expansion would be justified in regimes where hole excitations are  energetically more unfavorable than particle excitations, this corresponds to a low-energy density expansion.

This simple example illustrates the general idea of how form factor axioms, and therefore form factors themselves, are modified by the presence of the background, since the dependence on 
the backflow appears whenever a particle is crossed. Another important example is the annihilation-pole axiom, which is modified to
\beq
&&\hspace{-10pt}-{\rm i}{\rm Res}_{\theta_1=\theta_2}f_\rho^\mathcal{O}(\theta_1+\pi {\rm i},\theta_2,\theta_3,\dots,\theta_n)\nonumber\\
&& \hspace{10pt}=\left[1-R_\rho(\theta_2\vert \theta_3,\dots,\theta_n)S(\theta_2-\theta_3)\cdots S(\theta_2-\theta_n)\right]f_\rho^\mathcal{O}(\theta_3,\dots,\theta_n),\label{dressedannihilation}
\eeq
where
\beq
R_\rho(\theta_n\vert \theta_1,\dots,\theta_n)=\prod_{i=1}^n R_\rho(\theta_n\vert \theta_i).
\eeq

\subsection{Non-relativistic theories}
Thermodynamic form factors have been computed in the case of the non-relativistic Lieb-Liniger model in \cite{de2015density}, using a different strategy. The starting point is an exact expression for the general form factor with an arbitrary number of particles (\ref{generalformll}). The challenge is then to take the infinite volume limit of this expression. We will not discuss the details of this lengthy derivation here, which in essence amounts to reexpress Eq. (\ref{generalformll}), in the thermodynamic limit, in terms of a representative state and particle-hole excitations on top of this state. 

The density-operator form factor presented in \cite{de2015density} is a very cumbersome expression which will not be explored in detail. However, the form factor simplifies dramatically in the low density regime ($n(\lambda)\approx0$). In this limit, the form factor with $n$ particle-hole pair excitations is given by
\beq
&&\vert\langle \rho\vert \hat{\rho}(0)\vert \bar{\rho};h_1,\dots,h_n,p_1,\dots,p_n\rangle\vert=\frac{c}{2}\Big[\prod_{k=1}^n\sum_{l=1}^n(\theta(p_l-h_k)-\theta(h_l-h_k))\Big]\nonumber\\
&&\times\prod_{i,j}^n\Big[\frac{(p_i-h_i+{\rm i}c)^2}{(h_{ij}+{\rm i}c)(p_{ij}+{\rm i}c)}\Big]^{1/2}\frac{\prod_{i<j=1}^n h_{ij}p_{ij}}{\prod_{i,j}(p_i-h_j)}\det\big(\{\delta_{ij}+W(h_i,h_j)\}_{i,j=1}^{n}\big).
\eeq
with $\theta(\lambda)=2\arctan(\lambda/c)$ the two-particle phase shift, and $h_{ij}=h_i-h_j$, and $p_{ij}=p_i-p_j$. Here $\{h\}$ and $\{p\}$ are momenta of  holes and particles, respectively, and
\beq
W(\lambda,\mu)=b(\lambda)\left(\varphi_\mathrm{LL}(\lambda-\mu)-\frac{2}{c}\right)+\mathcal{O}(n(\lambda))
\eeq
 and
\beq
b(\lambda)& = &-\frac{\sin[\pi n(\lambda) \mathcal{F}(\lambda\vert \{h,p\})]}{2\pi n(\lambda)\mathcal{F}(\lambda\vert \{h,p\})\sin[\pi \mathcal{F}(\lambda\vert \{h,p\})]}\nonumber\\
&&\times\frac{\prod_{k=1}^n \mu_k^+-\lambda}{\prod_{k=1,\mu_k^-\neq \lambda}^{n}\mu_k^--\lambda}\prod_{k=1}^n\left(\frac{\varphi_\mathrm{LL}(\mu_k^+-\lambda)}{\varphi_\mathrm{LL}(\mu_k^--\lambda)}\right)^{\frac{1}{2}}\n
&&\times\exp\left[\frac{c}{2}\mathbf{PV}\int \dd\mu\frac{n(\mu)\mathcal{F}(\lambda\vert \{h,p\})\varphi_\mathrm{LL}(\lambda-\mu)}{\lambda-\mu}\right],\nonumber\\
\eeq
 where $\varphi_\mathrm{LL}(\lambda) = 2c/(\lambda^2+c^2)$, and $\mathcal{F}(\lambda\vert \{h,p\})$ is a sum of the backflow function (\ref{backflow}) defined by
 \begin{equation}
     \mathcal{F}(\lambda\vert \{h,p\})=\sum_{j=1}^n\left(\mathcal{F}(\lambda,h_j)-\mathcal{F}(\lambda,p_j)\right).
 \end{equation}

\subsection{Quantum spin chains}
So far we have been operating under the assumption that the thermodynamic two-point function can be approximated using only a few form factors with a small number of excitations on top of the thermodynamic background. This section on quantum spin chains is included so as to show that such an assumption may sometimes fail. The failure can be traced to the semi-local nature of kink-like excitations in the TFIC. To this end we focus on correlators of the operator $\sigma^x$.

The computation of thermodynamic two-point functions for the TFIC has been carried out in \cite{granet2020finite}, in the limit of large space-time separation between the operators, with fixed ratio $j/t=l$, using a finite-volume regularization. The idea is to consider two-point functions for a finite volume state $\vert q_1,\dots,q_N\rangle$, where $N$ is large and the rapidities are distributed according to $\rho(q)$. The finite-volume two-point function can then be computed term by term, using the finite volume form factors (\ref{isingformgeneral}), as
\beq
\chi(l,t)= \frac{\langle \rho\vert \sigma_{lt,t}^x\sigma^x_{0,0}\vert\rho\rangle}{\langle \rho\vert\rho\rangle}&=&\lim_{L,N\to\infty}\sum_{M=0}^{\infty}\frac{1}{M!}\sum_{p_1,\dots,p_M,q_1,\dots,q_N}\vert \langle p_1,\dots,p_M\vert \sigma_l^x\vert q_1,\dots,q_N\rangle\vert^2\nonumber\\
&&\times \mathrm{e}^{{\rm i}t(E(\{q\})-E(\{p\}))+{\rm i}lt(P(\{p\})-P(\{q\}))}.
\eeq
The form factor expansion assumes that  in the large $t$ limit the correlator is dominated by  form factors where $\{p\}\approx\{q\}$, up to a few particle and hole excitations. This way one is able to approximate the correlator by just including the first few dominant form factors. Such conjectures seem to be in contradiction with the main result in \cite{granet2020finite}. The long-distance correlator of TFIC involves an infinite number of contributions with arbitrary number of particles and holes. Rather than delving into the detailed derivation in \cite{granet2020finite}, we will focus on the physical arguments why this happens for the particular example.  

As conventional assumption, leading contributions to the correlator are those corresponding to a small number of particle and hole excitations and form factors with higher number of excitations adding subleading corrections at long distances. The leading term in the expansion would come from the one-particle-hole pair form factor,
\beq
\xi^{1ph}(l,t) = \lim_{L,N\to\infty}\sum_{p,h}\vert\langle q_1,\dots,q_N,p,h\vert \sigma_0^c\vert q_1,\dots,q_N\rangle\vert^2\,\exp\left({{\rm i}t(E(p)-E(h))+{\rm i}lt(p-h)}\right).\nonumber\\
\eeq
 In the infinite volume limit, the sum over values of $p$ and $h$ become an integral over momenta. For large values of  $t$, this term is expected to decay polynomially, which follows from performing a stationary phase approximation. As  a higher number of particles and holes are introduced, there will be more integrals over momenta which, by a stationary phase approximation, should be  subleading at large distances. This reasoning, while applicable to our previous examples, fails in our context, since particles are semi-local relative to $\sigma^x$. As discussed, this implies that the one-particle-hole-pair form factor has an annihilation pole. Integrating over momenta, there will be a non-trivial contribution arising from the region around this pole, which has no decay with distance, as assumed in the stationary phase approximation. The same argument holds for higher form factors.

The remaining  computation is a matter of identifying the leading long-distance terms for each form factor term. The computation in \cite{granet2020finite} is performed at a fixed, time-like value of $l=j/t$, at large $t$.  Generically, a form factor with $n$ particles and holes results in a leading contribution of order as  $t^n$. All these leading terms can actually be resummed within a low density expansion (in powers of $\rho(q)$, into an exponential (to order $\rho(q)^2$)),
\beq
\chi(l,t)\approx C\exp\left(-2\int_{-\pi}^\pi\dd x\rho(x)(1+2\pi \rho(x))\vert t(\varepsilon^\prime(x)-l)\vert \right),\nonumber
\eeq
where
\beq
C=\xi\exp\left(-2\int_{-\pi}^\pi\int_{-\pi}^\pi\dd x\dd y\frac{\rho(y)\rho^\prime(x)}{\tan\left(\frac{x-y}{2}\right)}\right).
\eeq

As we will see in the following section, a low order thermodynamic form factor expansion generally leads to expressions for correlation functions which match predictions from GHD. Such results do not apply to semi-local operators as those presented in this sections, as it is not sensible to truncate the form factor expansion to just a low number of particles and holes.

One intriguing possibility is to establish a general connection between form factor expansions of semi-local operators and the results from \cite{doyon2019fluctuations}, which merge GHD and large-deviation theory techniques to arrive to correlation functions of twist fields (which are characterized by their semilocality), which display similar exponential behavior at long distances.

\subsection{Challenges for independent verification of thermodynamic form factors}

The thermodynamic form factor formalism provides new tools for computing correlation functions that are inaccessible to other methods. This means that it is a challenge to provide an independent verification that the predictions from this formalism are correct.

Take for instance the case of standard ground-state form factors of QFT. Predictions from the form factor bootstrap have often been verified numerically through the truncated conformal spectrum approach (TCSA) \cite{yurov1990truncated}. The TCSA is a method for numerically diagonalizing the QFT Hamiltonian at finite volume, which allows one to easily obtained the low-lying state spectrum, and form factors between these states, and approach which has been used to successfully validate results from the form factor bootstap.

There are challenges to extend the TCSA to be able to compute {\it thermodynamic} form factors. The TCSA is naturally suitable for finding ground state and low-lying excitations, while to test thermodynamic bootstrap predictions, we need access to highly excited states with a finite energy density. 

There has been some progress in this direction in the context of quantum quenches. In \cite{kukuljan2018correlation} a quantum quench of the sine Gordon model was studied using TCSA, where the initial state is a low-lying eigenstate of a pre-quench Hamiltonian (which is thus accessible to TCSA). This allows a numerical computation of non-equilibrium and GGE correlation functions. 

There is some challenge in connecting the bootstrap and TCSA quantum quench results, in that the simplest models to study in both cases are different. The sine Gordon QFT has non-diagonal scattering, which makes it inaccessible to the simplest version of the form factor bootstrap presented here, which can only be applied directly to diagonal scattering theories (such as sinh Gordon). On the other hand, some properties of sinh Gordon make it not amenable to study through the TCSA. At this point either the bootstrap program needs to be generalized for non-diagonal scattering or TCSA needs to be modified to be able to treat models like sinh gordon, before a quantitative comparison between methods can be made.

A similar TCSA quantum quench computation has been done for the Lieb Liniger model \cite{robinson2019computing}, where the numerical convergence of TCSA was improved optimizing the truncation scheme used.  This has shown promising resuts, in allowing for the computation of non-equilibrium and GGE observables. For a small number of particles, this method was found to be in excellent agreement with coordinate Bethe ansatz calcuations \cite{zill2016coordinate}.

There are still some computational challenges left to overcome to make a proper comparison with thermodynamic form factors. For instance, The form factors discussed in this review are defined on the thermodynamic limit,  at finite energy density. TCSA works by definition with a finite number of particles, and it is better suited to study smaller quenches, where a small amount of energy  is introduced. A careful extrapolation study to larger number of particles, while keeping a finite energy density is needed to be able to make a proper quantitative comparison with the thermodynamic form factors discussed here.

\section{Form factors and  GHD predictions}

At this point we have built the scheme for approximating space-time correlation functions of local operators on a thermodynamic background. These results can be compared to predictions from the GHD linearized relative to the background.

One of the simplest results arising from the GHD formalism is the charge density correlator 
$\langle q_i(x,t)q_j(0,0)\rangle_\rho^\mathrm{c}$, the superscript $\mathrm{c}$ standing for the connected part of the correlation function. The factor 
$q_j(0,0)$  modifies the GGE initially at the origin and observed is how such a  perturbation propagates in space-time.
Thus on the Euler scale, this quantity should be approximated by the solution of the linearized GHD relative to the background root density $\rho$, \cite{Doyon_2017},

\beq\label{2-corr}
\lim_{\epsilon \to 0} \epsilon^{-1} \langle q_i( \epsilon^{-1}x, \epsilon^{-1}t)q_j(0,0)\rangle_\rho^\mathrm{c} = \int_\mathbb{R} \dd \theta \rho(\theta) f(\theta) h_i^{\rm dr}(\theta) h^{\rm dr}_j(\theta) \delta(x - v^{\rm eff}(\theta)t).\label{chargetwopoint}
\eeq
The dressed charge density, $h^{\rm dr}_j(\theta)$, is defined as per \eqref{dressing}
and the factor $f(\theta)$ depends on the type of particles considered. In our examples, the particles are  mostly of fermionic type for which  $f(\theta)=1-n(\theta)$. 

Part of the prediction \eqref{2-corr} is the correct choice of the effective velocity $v^\mathrm{eff}(\theta)$ as a nonlinear functional of the root density $\rho$. These average currents will be discussed at length in the second part of our review, one concise definition being \eqref{veff}.

We now turn to the computation of (\ref{chargetwopoint}) by means of the form factor formalism. Considering only connected correlation functions amounts to disregard the zero-particle/hole term in the expansion, $\langle\rho\vert q_1\vert\rho\rangle\langle\rho\vert q_2\vert\rho\rangle$. By stationary-phase arguments, an expression as (\ref{chargetwopoint}) should result from the one-particle-hole pair form factor. Since the one-particle-hole pair form factor has no annihilation pole, at large times the double integral over rapidities has a highly oscillatory integrand, which yields the $\delta$ function. Similarly one can argue that terms with higher number of particle and hole pairs will contribute at higher orders of $t^{-1}$ at late times. On the other hand, form factors with an unequal number of particles and holes are  highly oscillatory in $x,t$ and thus vanish under small scale averaging.
 
From these considerations, in order to reproduce the correlation function (\ref{chargetwopoint}), the one-particle-hole form factor should be of the form 
\beq
\lim_{\theta_h\to\theta_p}\frac{\langle \rho\vert q\vert \rho;\theta_p,\theta_h\rangle}{\langle\rho\vert\rho\rangle}=h^{\rm dr}(\theta_p).\label{chargeprediction}
\eeq
The form (\ref{chargeprediction}) has been already verified for the density operator of the Lieb-Liniger model \cite{De_Nardis_2018}. The form factor derived in \cite{de2015density} was shown to reduce to expression (\ref{chargeprediction}) for one-particle-hole pair in the zero momentum limit, referring to the particle and hole having the same momentum\footnote{We point out that the definition of thermodynamic form factor used in \cite{de2015density,De_Nardis_2018} is slightly different from ours, in that they include an extra factor which is meant to account for the summation over all the different soft modes that lead to the same macroscopic state configuration. This amounts to multiplying the form factor with  a term related to the change in the entropy of the state due to the addition of particles and holes. In the context of the results presented here, this difference in definition is irrelevant, as that additional factor reduces to one in the zero-momentum limit of the particle and hole that we consider.}.

In the case of a relativistic QFT, one can reach much beyond \eqref{2-corr} by establishing the Euler scaling form for general local observables, not only charge densities.  
To do so, one has to define more precisely the necessary averaging over microscopic scales (which was not needed for the conserved charge operators).
The first step is to consider an Euler macro-point $(x,t)$ with $(x,t)$ of order $1$, which on the microscopic scale is $(\lambda x, \lambda t)$ with $\lambda$ dimensionless and large. Following \cite{10.21468/SciPostPhys.5.5.054},
one considers a space-time region whose size scales as $\lambda^\nu$, for some $0<\nu_0 <\nu<1$, around the macro-point as $\mathcal{N}_\lambda(x,t)=\{(y,s)|(y-\lambda x)^2+(s-\lambda t)^2<\lambda^{2\nu}\}$. The $M$-point Euler-scale correlation functions are constructed by first picking
$M$ distinct macro-points $(x_1,t_1),\dots,(x_M,t_M)$ and averaging the local operator over the respective associated region $\mathcal{N}_\lambda(x_j,t_j)$.
As formula one thus arrives at 
\beq
&&\langle\mathcal{O}_1(x_1,t_1)\dots\mathcal{O}_N(x_M,t_M)\rangle_\rho^{\rm Eul}\\
&&=\lim_{\lambda\to\infty}\lambda^{M-1} \vert\mathcal{N}_\lambda\vert^{-M}
\int_{\mathcal{N}_{\lambda}(x_1,t_1)} \dd y_1 \dd s_1\dots\int_{\mathcal{N}_{\lambda}(x_M,t_M)}\dd y_M \dd s_M\langle\mathcal{O}_1(y_1,s_1)\dots\mathcal{O}_N(y_M,s_M)
\rangle^\mathrm{c}_\rho. \nonumber
\eeq
For the moment we will focus on the homogeneous case.


 With this definition, a general formula for Euler-scale correlators in QFT was derived in \cite{10.21468/SciPostPhys.5.5.054} from the GHD formalism, as
\beq\label{genEul}
\langle\mathcal{O}_1(x,t)\mathcal{O}_2(0,0)\rangle_\rho^{\rm Eul}=\int_\mathbb{R} \dd\theta\rho(\theta)f(\theta) V^{\mathcal{O}_1}(\theta) V^{\mathcal{O}_2}(
\theta)\delta(x - v^{\rm eff}(\theta)t),\label{generaltwopoint}
\eeq
where
\beq
V^{\mathcal{O}}(\theta)=\sum_{k=0}^\infty\frac{1}{k!}\int_{\mathbb{R}^k}\prod_{j=1}^k\left(\frac{\dd\theta_j}{2\pi}n(\theta_j)\right)(2\pi \rho^\tot(\theta))^{-1}f_c^{\mathcal{O}}(\theta_1,\dots,\theta_n,\theta).\label{LMtwo}
\eeq

In the GHD approach, the identity \eqref{genEul} is argued by means of hydrodynamic projections. On the microscopic time scale the operator $\mathcal{O}_2(0,0)$ is projected onto the conserved charges. This particular linear combination then evolves on the Euler time scale according to the linearized GHD as in \eqref{2-corr}. At the final time $t$ inserted is another hydrodynamic projection, now for the operator $\mathcal{O}_1(x,0)$.

We note the similarity between $V^\mathcal{O}(\theta)$ and the Leclair-Mussardo formula (\ref{leclairmussardo}), where the difference is the presence in the connected form factors of an additional particle of rapidity $\theta$. Through similar arguments, we see the expression (\ref{generaltwopoint}) is recovered from the form factor formalism through the limit 
\begin{equation}
    \lim_{\kappa\to0}f^\mathcal{O}_\rho(\theta+\pi \rm i,\theta+\kappa)=2\pi \rho^\tot(\theta) {\it V}^\mathcal{O}(\theta).\label{thingtoprove}
\end{equation}
Expression (\ref{thingtoprove}) has been proven in \cite{cubero2020generalized} mainly by taking the thermodynamic limit of a form factor of the form 
\begin{align}
    &f^\mathcal{O}_\vartheta(\theta+\pi{\rm i},\theta+\kappa)\nonumber\\
&=\Big(\lim_{L,n\to\infty}\frac{f^\mathcal{O}(\theta+\pi {\rm i},\theta_n+\pi{\rm i},\dots,\theta_1+\pi{\rm i},\theta_1+\kappa_1,\dots,\theta_n+\kappa_n,\theta+\kappa)}{\det G(\theta_1,\dots,\theta_n)}\Big)
(1+\mathcal{O}(\kappa)),\label{limph}
\end{align}
The parameters $\kappa_j=\kappa \varphi(\theta_j,\theta)/(2\pi L\rho^\tot(\theta_j))$ are dictated by the backflow of the particle-hole pair on the $n$ background particles.
The form factor on the right hand side of (\ref{limph}) involves $2n+2$ excitations which approach the location of their annihilation poles as $\kappa\to0$. The structure of this form factor can then be analyzed through the annihilation pole axiom (\ref{dressedannihilation}). Using (\ref{dressedannihilation}) the form factor can be re-expressed as a sum over the connected parts of form factors with $2n, 2n-2, 2n-4,\dots$ particles. After taking the thermodynamic limit of this sum of terms, it was shown in \cite{cubero2020generalized} that this series reproduces exactly the Leclair-Mussardo-like expression (\ref{thingtoprove},\ref{LMtwo}).

We point out that some significant progress has been made recently in going beyond these first order hydrodynamic regime  two point functions, specifically for the Lieb-Liniger model by exploring special parameter regimes. In the low-density regime, exact expressions for the two-point function were found in \cite{Granet_2021}. In the weakly interacting (small $c$) limit, the first, order $1/t^2$ corrections to the two-point function were obtained in \cite{10.21468/SciPostPhys.9.6.082}.

We briefly mention that there has also been progress in recovering GHD predictions for spatially inhomogeneous thermodynamic backgrounds, using QFT form factors. One-point functions of a local operator were studied in \cite{cubero2020generalized} for an initial inhomogeneous state of the form (\ref{inhomo}),
which is a generalized Gibbs ensemble, but with spatially dependent chemical potentials, a standard GGE being recovered for $\beta^i(\epsilon x)=\beta^i$.  A small value of $\epsilon$ ensures that these chemical potentials vary slowly with $x$.
The method in \cite{cubero2020generalized} consists in exploiting such slow spatial variation to approximate the true $\beta^i$'s by  functions piecewise constant over large hydrodynamic cells of size $l$, where $L/l\sim L$, yet $l$ is much larger than any length scale of the QFT.  In other words,
\beq
\int_\mathbb{R} \dd y\sum_{i=0}^\infty\beta^i(\epsilon y)q_i(y)\approx\sum_K\int_{K-\frac{l}{2}}^{K+\frac{l}{2}}\dd y\sum_{i=0}^\infty\beta_K^iq_i(y),
\eeq
where the index $K$ denotes the cell number.
Furthermore, one can define an arbitrary {\it homogeneous} set of chemical potentials, $\{\bar{\beta}\}$, and a set of operators which have support only on the cell $K$,
\beq
O_K=\exp\Big[\int_{K-\frac{l}{2}}^{K+\frac{l}{2}}\dd y\sum_{i=0}^\infty\bar{\beta}^iq_i(y)\big]=\exp\Big[-\int_{K-\frac{l}{2}}^{K+\frac{l}{2}}
\dd y\sum_{i=0}^\infty\beta_K^iq_i(y)\Big].
\eeq
The purpose of defining these quantities is that we can use them to re-express the one-point function in the inhomogeneous background, in a way that resembles a many-point function in a homogeneous background. The specific values of $\{\bar\beta\}$ do not need to be specified at this point, and we will use this arbitrariness later to fix the values that yield the simplest expression for the one-point function. As shown in \cite{cubero2020generalized}, for these operators, at large $l$ and $L$, the one-point function in the state (\ref{inhomo}) is approximated for large $t$ by
\beq
\langle\mathcal{O}(x,t)\rangle\textcolor{red}{\approx}\frac{{\rm Tr}\left[e^{-\sum_i\bar{\beta}^iQ_i}\mathcal{X}\left(\mathcal{O}(x,t)\prod_{K_-<K<K_+}O_K\right)\right]}{{\rm Tr}\left[e^{-\sum_i\bar{\beta}^iQ_i}\mathcal{X}\left(\prod_{K_-<K<K_+}O_K\right)\right]},\label{homogeneouscorrelator}
\eeq
where $\mathcal{X}$ stands for $x$-ordering of the operators and the range $K\in[K_-,K_+]$ are those cells which lie within the past light-cone of the point $x,t$.

Expression (\ref{homogeneouscorrelator}) is now a standard correlation function in a homogeneous thermodynamic background, defined by $\{\bar{\beta}\}$ which yet remain to be fixed. This correlator can therefore be evaluated using form factor expansions. Again the leading contributions arise from one-particle hole pair form factors.

These one-particle form factor contributions, however can be shown to vanish by making a a clever choice of the chemical potentials $\bar{\beta}^i(x,t), i = 0,\dots, \infty$, which are still free parameters we can adjust. After choosing the most convenient set of chemical potentials, the leading contribution at late times can be expressed in terms of only zero-particle (or expectation values) thermodynamic form factors. It is shown in  \cite{cubero2020generalized} that this convenient choice of chemical potentials is exactly the one corresponding to having $\bar{\beta}^i(x,t), i = 0,\dots, \infty,$  evolved according to GHD. 
One can therefore approximate for each $x,t$,
\beq\label{appro}
\langle\mathcal{O}(x,t)\rangle\simeq\frac{ \langle \rho_{\{\bar{\beta}\}_{x,t}}\vert\mathcal{O}(0,0)\vert\rho_{\{\bar{\beta}\}_{x,t}}\rangle}{ \langle \rho_{\{\bar{\beta}\}_{x,t}}\vert\rho_{\{\bar{\beta}\}_{x,t}}\rangle}
\eeq

As alluded to in the introductory Section \ref{ghdintro}, GHD relies on the fundamental assumption of propagation of local GGE.
This property is indeed confirmed through the approximation \eqref{appro}, including the specific effective velocity as in \eqref{veff}. \newpage
\noindent
{\Large{\textbf{Part II: Collision rate ansatz}}}

\section{Phenomenology of collision rate ansatz}

In Section \ref{ghdintro}  the effective velocity was introduced abstractly, while for GHD required is the functional dependence on $\rho$.
As discovered and argued in \cite{PhysRevX.6.041065,PhysRevLett.117.207201}, $v^\mathrm{eff}(\theta)$ is  in fact determined  as solution of the linear integral equation
\begin{equation}\label{veff}
    v^\mathrm{eff}(\theta)=\frac{E'(\theta)}{p'(\theta)}+\int_\mathbb{R}\mathrm{d}\theta'\frac{\varphi(\theta,\theta')}{p'(\theta)}\rho(\theta')\big(v^\mathrm{eff}(\vartheta')-v^\mathrm{eff}(\theta)\big).
\end{equation}
Here $E,p$ are energy and momentum of the considered integrable system and $\varphi(\theta,\theta')$ is the respective two-particle scattering shift. The appearance of $p'(\theta)$ in the integral is due to the change of variable from momentum to rapidity. At the moment Eq. \eqref{veff}  is regarded as an ansatz and the underlying physical reasoning will be explained right away.
More efforts are required to establish the precise connection to the GGE average current $\langle j_i(0)\rangle_\rho$. 
\begin{figure}[ht!]
\centering
    \includegraphics[width=0.5\linewidth]{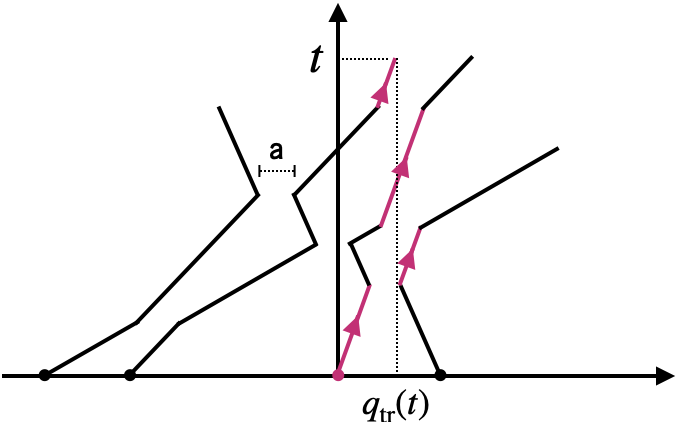}
    \caption{Dynamics of hard rods with a trajectory of the tracer quasi-particle indicated in magenta.}
    \label{rods}
   \end{figure}

To argue for the validity of \eqref{veff}, it is instructive to consider a system of point particles on the line 
 interacting through the hard core potential $V(x) = \infty$ for $|x| \leq \mathsf{a}/2$ and $V(x) = 0$ for $|x| > \mathsf{a}/2$. 
This system can be realized as a collection of hard rods with rod length  $\mathsf{a}$. GHD for hard rods has been studied on a mathematical level in \cite{Boldrighini1983} with Navier-Stokes corrections in \cite{Boldrighini1997}, see also \cite{spohn2021hydrodynamic}. Hard rods rattle back and forth between their neighbors. Quasi-particles are introduced by 
 following trajectories at constant velocity. A quasi-particle moves with given velocity, except for  jumps in position by  $\mathsf{a}$ to the right and by $-\mathsf{a}$ to the left. The sum of an arbitrary one-particle velocity function is conserved, hence the system is integrable. In a GGE velocities are independent with probability density function $f(v) \geq 0$, normalized as  $\int_\mathbb{R} \dd v f(v) = 1$.
 The positions are uniform under the hard core constraint. We prepare now an infinitely extended GGE and mark the quasi-particle 
 next to the origin as tracer particle. Its position, $q_\mathrm{tr}(t)$, at time $t$ is given by
 \begin{equation}\label{veffmicro}
  q_\mathrm{tr}(t)   =v t-C_\mathrm{l}(t)\mathsf{a}+C_\mathrm{r}(t)\mathsf{a}
 \end{equation}
 with $C_\mathrm{l}(t)$ the number of left and $C_\mathrm{r}(t)$  of right collisions up to time $t$. For hard rods these collisions are statistically independent. By a law of large numbers $q_\mathrm{tr}(t) \simeq v^\mathrm{eff}t + \mathcal{O}(\sqrt{t})$ for large $t$ with the effective velocity 
 \begin{equation}\label{veffhr}
 v^\mathrm{eff}(v) =  \frac{1}{1 - \mathsf{a} \rho}(v - \mathsf{a} \rho u),
\end{equation} 
 where $\rho$ is the particle density and $u$ the mean velocity of the hard rod fluid.
 
 To establish the connection with \eqref{veff}, we  set $E'(\theta) = \theta$, $p'(\theta) = 1$, and replace $\theta$ by $v$, $ \rho(\theta)$ by
 $\rho f(v)$. The scattering shift of hard rods is independent of incoming velocities  and given by  $- \mathsf{a}$. Thus the transcription 
 reads
 \begin{equation}\label{rodveff} 
     v^\mathrm{eff}(v)=v- \mathsf{a}\int_\mathbb{R} \dd w\rho f(w)\big(v^\mathrm{eff}(w) - v^\mathrm{eff}(v)\big).
 \end{equation}
 To confirm \eqref{rodveff}, needed is only the collision rate for the tracer quasi-particle.  $\rho h(v)$ is the thermodynamic background characteristic for the
 specific GGE. The tracer quasi-particle moves with bare velocity $v$. It is modified by collisions with background quasi-particles of 
 bare velocity $w$. The jump distance is  $- \mathsf{a}$ and the collision rate, i.e. number of collisions per unit time, is given by the last round bracket.   Indeed, the solution of the integral equation agrees with \eqref{veffhr}.
 
 For a general integrable system, the same kind of reasoning can be repeated. Now the two-particle scattering shift is velocity dependent 
 and collisions are no longer independent. Equation \eqref{veff} states that the effective velocity can still be computed as if incoming velocities are independent. This is why \eqref{veff} is called \textit{collision rate ansatz}.
 
 A merit of the ansatz lies in its wide and flexible applicability. Just to illustrate the case, suppose we want to write down GHD for the classical Calogero-Moser model, which is a one-dimensional fluid with  repulsive pair potential $(\sinh(x))^{-2}$ \cite{Calogero1971}. The two-particle scattering shift is known
 and $\rho(\theta)$ will be related to the density of states of the Lax matrix. With a good level of confidence, the current densities for the 
 Calogero-Moser model are still given by \eqref{veff}.  But a literal confirmation of \eqref{veff} will be difficult, the first obstacle being that already the notion of quasi-particle is vague. Therefore there must be a deeper reason for the validity of the collision rate ansatz, even more so since in \eqref{veff} any GGE is allowed.
 
 For a classical particle with energy-momentum relation $E(p)$, the velocity is defined by $v(p) = \partial_p E(p)$. If $E(p)$ is given parametrically as $E(\theta),p(\theta)$ then $v(\theta) = \partial_\theta E(\theta)/\partial_\theta p(\theta)$. As unique solution to the rate equation,
 one can show \cite{PhysRevLett.113.187203} that such a relation persists as
 \begin{equation}\label{veff1}
    v^\mathrm{eff}(\theta)=\frac{\partial_\theta E_{\mathrm{eff}}(\theta)}{\partial_\theta p_{\mathrm{eff}}(\theta)},
\end{equation}
where we recall that $E_\eff(\theta)$ and $p_\eff(\theta)$ are the excited energy and momentum that are incurred by adding a quasi-particle $\theta$ to a GGE background defined in \eqref{effectiveep}. With such a relation, the physics intuition
is shifted to quasi-particles as classical objects with an energy-momentum relation resulting from the underlying GGE. 
  
Before jumping into the rigorous derivation of the collision rate ansatz, let us introduce two further heuristic ways of understanding how the actual form of the effective velocity can be envisaged. First, consider the Lieb-Liniger $\delta$-Bose gas at zero temperature, which is a filled Fermi sea \cite{PhysRev.130.1605}. One then imposes a weak perturbation of the Fermi sea, which generates two ripples made of excitations having a common velocity, namely the Fermi velocity (or sound velocity) $v_\mathrm{F}(\theta_\mathrm{F})$ defined by
\begin{equation}\label{veffgs}
    v_\mathrm{F}(\theta_\mathrm{F})=\left.\frac{\partial E_{\mathrm{eff}}(\theta)}{\partial p_{\mathrm{eff}}(\theta)}\right|_{T=0,  \,\theta=\theta_\mathrm{F}},
\end{equation}
where $\theta_\mathrm{F}$ is the Fermi quasi-momentum. By taking the limit $T\to0$, the excitation boils down to the usual particle excitation of the Fermi sea. Just two ripples are generated, since the dispersion relation near the ground state is linear, i.e. $\omega(p)= v_\mathrm{F}|p|$ for the Lieb-Liniger gas. In other words, the ground state of the Lieb-Liniger model is a Luttinger liquid. Hence the deviation of the root density $\rho(\theta)\simeq \rho(\theta_\mathrm{F})+\delta\rho(\theta)$ satisfies the continuity equation 
\begin{equation}
    \partial_t\delta\rho(\theta)+\mathrm{sgn}(\theta)v_\mathrm{F}\partial_x\delta\rho(\theta)=0,
\end{equation}
where $\mathrm{sgn}(\theta)$ reflects the double branch of the dispersion relation. Next let us consider a weak perturbation of the Lieb-Liniger gas at finite temperatures. Now the dispersion relation for the excitations on top of thermal equilibrium is highly nonlinear and, in principle, each quasi-particle with quasi-momentum $\theta$ is carried with a distinct velocity. A natural generalization of $v_\mathrm{F}$ in such a situation is to view the velocity of excitations on top of some thermal equilibrium rather than the ground state, which simply amounts to  \eqref{veffgs} without taking the $T=0$ limit, i.e. \eqref{veff1}. 
With this velocity, it is then natural to expect that the root density $\rho(\theta)$ is transported by the continuity equation
\begin{equation}
    \partial_t\delta\rho(\theta)+v^\mathrm{eff}(\theta)\partial_x\delta\rho(\theta)=0,
\end{equation}
which happens to be the linearized GHD equation. Therefore we just demonstrated that, even without knowing anything about GHD, we can argue how the root density should be transported using the analogy to a Luttinger liquid. GHD however tells us that much more is true: even far from equilibrium, still requiring slow variation, quasi-particles are transported with velocity $v^\mathrm{eff}(\theta)$, which is now promoted to be space-time dependent on the hydrodynamic scale. 

A second approach seems to work only for relativistic systems and leads to the anticipated expression for the average current, hence also for the effective velocity. Let us invoke the crossing symmetry of relativistic systems, which states that the role of charge densities $q_i$ and current densities $j_i$ can be exchanged when looking at the system from the cross channel. At the level of quasi-particles, such mapping is accomplished by $\theta\mapsto\theta- \pi\mathrm{i}/2$, which yields $(E,p)\mapsto(\mathrm{i} p,-\mathrm{i} E)$. This suggests that, upon Wick rotation, the average charge density with the cross channel should be identified with the average current density in the actual channel. Implementing this idea in \eqref{chargeaverage}, one concludes that the average current reads \cite{PhysRevX.6.041065}
\begin{equation}\label{currentaverage}
    \langle j_i\rangle=\int\frac{\mathrm{d}\theta}{2\pi}E'(\theta)n(\theta)h^\mathrm{dr}_i(\theta)=\int\mathrm{d}\theta\rho(\theta)v^\mathrm{eff}(\theta)h_i(\theta)
\end{equation}
 with the effective velocity $v^\mathrm{eff}(\theta)$ given by \eqref{veff}.

As seen so far, the validity of the collision rate ansatz \eqref{veff} can be argued making use of rather diverse reasonings. A more down-to-earth approach can actually be found in one of the very first GHD papers \cite{PhysRevX.6.041065}, where the issue was approached through form factor expansions. Later a first full proof of the ansatz was accomplished by using graph theoretical methods to fill a gap in the prior work. Although 
in its initial form this proof is applicable only to relativistic systems, the same strategy is valid also for integrable spin chains \cite{PhysRevX.10.011054}, for which exact finite volume matrix elements of the current operators are obtained. In addition to these proofs, two novel derivations have  become available recently. One of them is based on long-range deformations of integrable lattice systems \cite{10.21468/SciPostPhys.8.2.016}, while the second one relies on the symmetry of the charge-current susceptibility matrix together with  the existence of a self-conserved current \cite{10.21468/SciPostPhys.9.3.040}. In the latter proof, it was observed that the availability of a self-conserved current is closely related to the existence of a boost operator, which also appears in the former approach. Even more recently the algebraic construction of the current operators has been accomplished, thereby serving as a new rigorous derivation of the collision rate ansatz \cite{PhysRevLett.125.070602}. In the following sections, we will explain in detail two methods, namely form factor expansions and  self-conserved current. The other approaches are covered in a companion review in the same volume \cite{borsi2021current}.

The studies mentioned deal  with quantum integrable systems. However, it is important to remark that actually the very first appearance of the collision rate conjecture dates back to half a century ago in the context of the classical soliton gas \cite{WOS:A1971K293500013}. For the history of kinetic approaches to soliton gases and integrable turbulence, see the review \cite{el2021soliton} in the same volume. More recently the ansatz has also been applied to other classical hamiltonian
integrable models, foremost the classical Toda lattice \cite{doi:10.1063/1.5096892,Spohn2019}. For this model,  the method of a self-conserved current is particularly transparent, see  \cite{PhysRevE.101.060103} with earlier numerical confirmations \cite{Cao_2019}.

\section{Rate ansatz from form factor expansion}
Our first proof relies on the standard zero density form factor expansion, which does not necessitate the use of thermodynamic form factors \cite{10.21468/SciPostPhys.6.2.023}. This point is particularly important because it allows us to carry out first-principle computations solely based on the axioms of form factors without any further assumptions or approximations. Note also that in Sect. 6 the collision rate ansatz, i.e. the functional form of $v^\eff(\theta)$, was actually justified through the approximation by a local GGE, as well as using the thermodynamic form factor expansion. In this section, the proof does not require any underlying assumption, such as local GGE, and the main idea turns out to be  universal (the same combinatorial idea has been used in \cite{PhysRevX.10.011054} in the context of Heisenberg spin chains).

Let us outline how the original proof works. The strategy is actually rather simple: one starts with the connected form factor of the charge density \eqref{chargeconnff}, from which, the symmetric form factor of the charge density is obtained, by using the formula \eqref{symmconn}. The symmetric form factor of the charge density can then be easily transformed into the one of the current via the continuity equation, $\p_t q_i+\p_x j_i=0$. Once more using \eqref{symmconn}, but this time backwards,  the connected form factor of the current is obtained. Plugging into the LeClair-Mussardo formula, we finally obtain the current average \eqref{currentaverage} with $v^\eff(\theta)$ of \eqref{veff}, thereby establishing the collision rate ansatz. As a by-product, by the same method the finite-volume matrix element of the current can be derived. Note that the finite-volume matrix element of the current was obtained for the first time in \cite{PhysRevX.10.011054} for the XXZ spin-1/2 chain.


As crucial observation, facilitating the proof, the principal minor $L(\alpha|\alpha)$ that appeared in \eqref{symmconn} bears a graph theoretical interpretation. To appreciate such an interpretation, we recall some notions from graph theory before presenting the proof.

\begin{figure}[ht!]
\centering
    \includegraphics[width=0.8\linewidth]{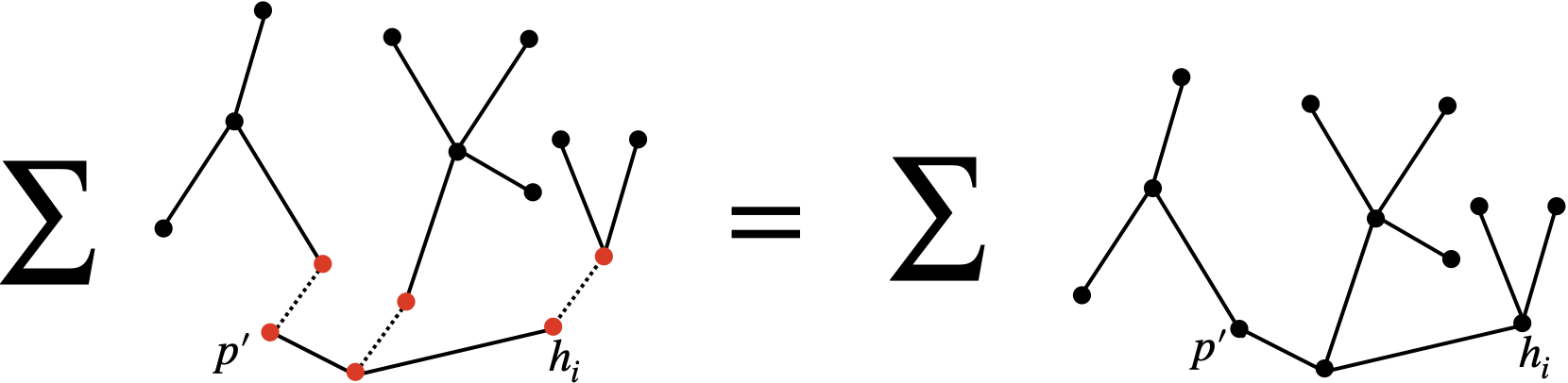}
    \caption{Merger of forests and spines. They are sewn together along the roots (red vertices) connected by the dashed lines.}
    \label{generating}
   \end{figure}

First, a {\it graph} is a set of vertices and edges $(V,E)$. If each edge has a direction, the graph is called a directed graph. If some numerical value is assigned to each edge of a graph, then the graph is called a weighted graph. In our argument we shall primarily deal with {\it trees}, which are connected graphs without cycles. A {\it forest} is then a set of trees. If all the available vertices in a given graph are covered by a tree or a forest, then the graph is called a spanning tree/forest. A diverging tree is a directed tree that contains directed paths from one marked vertex called {\it root} to all the other vertices. A diverging forest is then a collection of divergent trees.  
With these notions, we are ready to spell out the theorem central in the proof. Let $\alpha$ be a subset of the set $\{1,\cdots,n\}$ of integers. Then the {\it matrix tree theorem} states that  $L(\alpha|\alpha)$ can be written as the following sum of forests \cite{CHAIKEN1978377}
\begin{equation}
    L(\alpha|\alpha)=\sum_{F\in\mathcal{F}_\alpha}\prod_{\ell_{jk}\in E[F]}\varphi_{j,k},
\end{equation}
where $\mathcal{F}_\alpha$ is the set of spanning undirected forests with $n$ vertices such that  each tree contains exactly one vertex from $\alpha$,  which we call $\alpha$-vertex, and $E[F]$ is the set of edges of the forest $F$. Such a combinatoric reinterpretation of the otherwise elusive object $L(\alpha|\alpha)$ turns out to be extremely effective. Let us examine \eqref{symmconn}. An important observation is that $f_c^q(\{\theta_i\}_{i\in\alpha})$ is a sum of spines with vertices that are also labeled by the subset $\alpha$. This suggests that the right hand side in \eqref{symmconn} actually  decorates the spines by attaching the $\alpha$-vertices of the trees to the vertices of the spines, see Fig.\,\ref{generating}. As a result one obtains the spanning trees consisting of $n$ vertices with $p'$ and $h_i$ inserted at two arbitrary positions in each tree. Note that double counting of the same tree does not happen because each tree is now labeled. Therefore the symmetric form factor can be written as
\begin{align}\label{symmq}
    f_\mathrm{s}^{q_i}(\theta_1,\dots,\theta_n)&=\sum_{\substack{\alpha \subset \{1,\dots,n\}\\
\alpha\neq\varnothing}}L(\alpha|\alpha)f^{q_i}_\mathrm{c}(\{\theta_i\}_{i\in\alpha})\n
&=\sum_{\substack{\alpha \subset \{1,\dots,n\}\\
\alpha\neq\varnothing}}\sum_{F\in\mathcal{F}_\alpha}\prod_{\ell_{jk}\in E[F]}\varphi_{j,k}\left(h_i(\theta_{i_1})\varphi_{i_1,i_2}\varphi_{i_2,i_3}\cdots\varphi_{i_{|\alpha|}-1,i_{|\alpha|}}p'(\theta_{i_{|\alpha|}}) + \mathrm{perm.}\right)\n
&=\sum_{\substack{\alpha \subset \{1,\dots,n\}\\
\alpha\neq\varnothing}}\left(\parbox{25em}{sum of spanning trees with $p'$ and $h_i$ being inserted at two arbitrary vertices that are contained in $\alpha$.}\right) \n
&=\sum_{j=1}^nh_i(\theta_j)\sum_{k=1}^np'(\theta_k)\sum_{T\in\mathcal{T}}\prod_{\ell_{jk}\in E[T]}\varphi_{j,k},
\end{align}
where $i_1,i_2,\cdots,i_{|\alpha|}\in\alpha$ with $|\alpha|$ the cardinality of the set $\alpha$, and $\mathcal{T}$ is the set of the spanning trees.
Next, it is a simple matter to note that the continuity equations at the operator level, $\p_t q_i+\p_x j_i=0$, allows us to relate the symmetric form factor of $q_i$ and $j_i$ as \cite{10.21468/SciPostPhys.6.2.023}, 
\begin{equation}\label{symmqj}
   f_\mathrm{s}^{j_i}(\theta_1,\dots,\theta_n)=\frac{\sum_{k}E'(\theta_k)}{\sum_{k}p'(\theta_k)}f_\mathrm{s}^{q_i}(\theta_1,\dots,\theta_n),
\end{equation}
which, together with \eqref{symmq}, implies
\begin{equation}\label{symmj}
    f_\mathrm{s}^{j_i}(\theta_1,\dots,\theta_n)=\sum_{j=1}^nh_i(\theta_j)\sum_{k=1}^nE'(\theta_k)\sum_{T\in\mathcal{T}}\prod_{\ell_{jk}\in E[T]}\varphi_{j,k}.
\end{equation}
The symmetric form factor of the current at can be transform back to its connected form factor by invoking \eqref{symmconn}. More precisely, one can extract a unique spine that connects $E'$ and $h_i$ in each tree due to the connectedness and the absence of cycles in each tree. As a net result,  
the same decomposition of the symmetric form factor into the sum of spanning trees and spines is obtained in the same form  as in $f_\mathrm{s}^{q_i}(\theta_1,\dots,\theta_n)$, except that $p'$ is now replaced by $E'$.  Thereby the connected form factor of the current $j_i$ is given by
\begin{equation}
    f_\mathrm{c}^{j_i}(\theta_1,\cdots,\theta_n)=h(\theta_1)\varphi_{1,2}\cdots\varphi_{n-1,n}E'(\theta_n) + {\rm perm.}\,.
\end{equation}
From the LeClair-Mussardo formula \eqref{LM} one then derives the average current  formula
\begin{equation}
       \langle j_i\rangle=\int\frac{\dd\theta}{2\pi}(E')^\dr(\theta) n(\theta)h_i(\theta)=\int\dd\theta\rho(\theta)v^\eff(\theta)h_i(\theta).
   \end{equation}

We turn to the finite-volume matrix element of the current operator $j$, which is valid up to  exponentially decaying terms in $L$. In fact a neat expression results from
\begin{equation}\label{finvolcurr}
    {}_L\langle\theta_n,\dots,\theta_1|j_i|\theta_1,\dots,\theta_n\rangle_L=\det G(\theta_1,\dots,\theta_n)\sum_{j,k=1}^nE'(\theta_j)G^{-1}_{jk}h_i(\theta_k)+\mathcal{O}(e^{-\mu L}),
\end{equation}
where $G^{-1}$ is the inverse of the $n\times n$ Gaudin matrix. To prove \eqref{finvolcurr}, let us first recall that, according to the Pozsgay-Tak\'acs formula \eqref{diag}, the finite-volume diagonal matrix element of the current operator $j_i$ can be written in terms of its connected form factor as
\begin{equation}\label{currentdiag}
     {}_L\langle\theta_n,\dots,\theta_1|j_i|\theta_1,\dots,\theta_n\rangle_L=\sum_{\alpha\subset\{1,\cdots,n\}}f_\mathrm{c}^{j_i}(\{\theta_i\}_{i\in \alpha})G(\alpha|\alpha)+\mathcal{O}(e^{-\mu L}).
\end{equation}
Since $G_{jk}=Lp'_j\delta_{jk}+L_{jk}$ where $p'_j=p'(\theta_j)$ and hence
\begin{equation}
    G(\alpha|\alpha)=\sum_{I\subset\bar{\alpha}}L(\alpha\cup I|\alpha\cup I)\prod_{j\in I}Lp'_j=\sum_{I\subset\bar{\alpha}}\sum_{F\in\mathcal{F}_{\alpha\cup I}}\prod_{j\in I}Lp'_j\prod_{\ell_{jk}\in E[F]}\varphi_{j,k},
\end{equation}
 the right hand side in \eqref{currentdiag} yields the summation over all forests, each of which contains a tree in which $E'$ is inserted at one of the vertices and $h$ is inserted at another vertex, and other trees possess exactly one vertex with weight $Lp'$. This implies that all we have to do is show that $G^{-1}_{jk}$ generates all forests each tree of which contains one vertex with weight $Lp'$ except for one tree which is obtained by decoration of a spine having a vertex $j$ at one end and  $i$ at the other one. Namely, 
\begin{equation}\label{ginv}
    G^{-1}_{jk}=\frac{1}{\det G}\sum_{I\subset\{1,\cdots,n\}\backslash\{j,k\}}\sum_{F\in\mathcal{F}_{I,jk}}\prod_{m\in I}Lp'_m\prod_{\ell_{lk}\in E[F]}\varphi_{l,k},
\end{equation}
where $\mathcal{F}_{I,jk}$ refers to the set of forests each tree of which has one vertex from $I$ except for one tree which contains $i$ and $j$ at two arbitrary positions. Note that $\det G$ in the denominator is needed to cancel out another $\det G$ that appears in \eqref{finvolcurr}. This expression, when plugged in the right hand side of \eqref{finvolcurr}, generates the desired forest expansion.

To confirm, let us start with rewriting the Gaudin matrix $G_{jk}$ as $G_{jk}=Lp'_k\Tilde{G}_{jk}$, where $\Tilde{G}_{jk}$ is given by
\begin{equation}
    \Tilde{G}_{jk}=\delta_{jk}+\delta_{jk}\sum_{l=1,l\neq j}^n\Tilde{\varphi}_{j,l}-(1-\delta_{jk})\Tilde{\varphi}_{j,k},\quad \Tilde{\varphi}_{j,k}=\frac{1}{Lp'_k}\varphi_{j,k}.
\end{equation}
Then, according to Theorem 3 in \cite{2006math......2070C}, one has
\begin{equation}\label{theorem3}
     LG^{-1}_{jk}p'_j=\Tilde{G}^{-1}_{jk}=\frac{1}{\det\tilde{G}}\sum_{F\in\mathcal{F}^\mathrm{d}_{j\rightsquigarrow k}}\prod_{\ell_{lk}\in E[F]}\Tilde{\varphi}_{l,k},
\end{equation}
where $\mathcal{F}^\mathrm{d}_{j\rightsquigarrow k}$ is the set of directed forests such that the vertices labeled by $j$ and $k$ belong to the same tree, diverging from $j$. By pulling out the factor $Lp'_k$ both from $\Tilde{\varphi}_{j,k}$ and $\Tilde{G}_{jk}$, we notice that these factors cancel out except those assigned to the roots including $k$. Therefore \eqref{theorem3} can be recast into 
\begin{equation}
  LG^{-1}_{jk}p'_j= \frac{1}{\det G}\sum_{I\subset\{1,\cdots,n\}\backslash\{j,k\}}\sum_{F\in\mathcal{F}_{I,jk}}\prod_{m\in I\cup\{j\}}Lp'_m\prod_{\ell_{lk}\in E[F]}\varphi_{l,k},
\end{equation}
which is equivalent to \eqref{ginv} after dividing both sides by $Lp'_j$. Thereby  the proof is completed.  As mentioned above the same result has also been obtained for the XXZ spin-1/2 chain by explicitly proving a relation similar to \eqref{ginv} \cite{PhysRevX.10.011054}.

 \section{Rate ansatz from a self-conserved current}
 \subsection{Symmetry of the charge-current susceptibility matrix}

The collision rate ansatz can be established by appealing to the symmetry of charge-current susceptibility matrix. A related approach is based on the algebra satisfied by the conserved charges and the boost operator. 

To start, let us recall the definitions of the static charge-charge and charge-current susceptibility matrices which 
are defined by \cite{SciPostPhys.3.6.039} 
\begin{align}
    C_{ij}=&\sum_{x\in\mathbb{Z}}\langle q_i(x,0)q_j(0,0)\rangle^\mathrm{c}_\rho= -\frac{\p}{\p\beta^i}\langle q_j\rangle_\rho,\\
    B_{ij}=&\sum_{x\in\mathbb{Z}} \langle q_i(x,0)j_j(0,0)\rangle^\mathrm{c}_\rho=  -\frac{\p}{\p\beta^i}\langle j_j\rangle_\rho.
\end{align}
These expressions refer to the infinitely extended system and, as before, the superscript $\langle\cdot\rangle^\mathrm{c}$ denotes truncated correlations, $\langle ab\rangle^\mathrm{c}_\rho =\langle ab\rangle_\rho-\langle a\rangle_\rho\langle b\rangle_\rho$. Somewhat surprisingly, not only the matrix $C_{ij}$ but also $B_{ij}$ turns out to be symmetric. Indeed, while the symmetry of the $B$ matrix has been known in the context of hydrodynamics \cite{Tth2003,Grisi2011,Spohn2014,PhysRevX.6.041065,SciPostPhys.3.6.039,10.21468/SciPostPhys.6.4.049}, the understanding on its full implications have been largely missing. This symmetry was noticed to hold in interacting particle systems with many conservation laws in \cite{Tth2003}, and later proved for stochastic interacting particle systems in \cite{Grisi2011}. In the context of quantum many-body systems, the symmetry  was rediscovered in \cite{PhysRevX.6.041065} and exploited to argue that, in analogy to the free energy, there must be a functional whose derivatives are the average currents. The symmetry actually holds even without spatial integration by assuming translation invariance as well as a strong clustering property. Precise conditions for the symmetry to hold, both with and without spatial integration, were established in \cite{10.21468/SciPostPhys.6.6.068}.

Let us briefly recall the conditions required for the $B$ matrix to be symmetric. As established in \cite{10.21468/SciPostPhys.6.6.068}  for spin chains,  time-stationarity of the density operator and continuity equations imply
\begin{equation}\label{bmat1}
    \langle j_i(0)Q_j\rangle^\mathrm{c}_L-\langle Q_ij_j(0)\rangle^\mathrm{c}_L=L\left(\mathcal{B}_{ij;L}(\lfloor L/2\rfloor+1,0)-\Tilde{\mathcal{B}}_{ij;L}(\lfloor L/2\rfloor,0)\right).
\end{equation}
Here the finite volume correlators are
\begin{align}
    \mathcal{B}_{ij;L}(x,t)&=\frac{1}{L}\sum_{y\in\mathcal{S}}\langle j_i(y,t)q_j(y+x,0)\rangle^\mathrm{c}_L,\n
     \Tilde{\mathcal{B}}_{ij;L}(x,t)&=\frac{1}{L}\sum_{y\in\mathcal{S}}\langle q_i(y,t)j_j(y+x,0)\rangle^\mathrm{c}_L,
\end{align}
the subscript $L$ indicating a system on a finite ring $\mathcal{S}=\{1,\dots,L\}$. 
Assuming that the right hand side of \eqref{bmat1} vanishes when $L\to\infty$ and noting that $\langle\cdot\rangle_\infty=\langle\cdot\rangle_\rho$ by definition, one can establish that $\langle j_i(0)Q_j\rangle^\mathrm{c}_\rho=\langle Q_ij_j(0)\rangle^\mathrm{c}_\rho$, i.e. $B_{ij}=B_{ji}$. 
In the above argument spatial translation invariance is never used. Translation invariance, together with the mentioned decay of correlation functions, however helps to establish a stronger result.  The symmetry holds even without the summation over the lattice, i.e.
\begin{equation}
    \langle j_i(0,0)q_j(x,t)\rangle^\mathrm{c}_\rho=\langle q_i(x,t)
j_j(0,0)\rangle^\mathrm{c}_\rho.
\end{equation}
In fact this symmetry leads to a peculiar factorization property of the combination of correlation functions
\begin{equation}
    \langle j_i(0,0)q_j(x,t)\rangle_\rho-\langle q_i(x,t)j_j(0,0)\rangle_\rho=\langle j_i\rangle_\rho\langle q_j\rangle_\rho-\langle j_j\rangle_\rho\langle q_i\rangle_\rho,
\end{equation}
which turns out to be at the root of the solvability of $T\Bar{T}$-deformed theories \cite{Jiang_2021}, or more generally Smirnov-Zamolodchikov type of deformation \cite{SMIRNOV2017363}. One of the important consequences of this symmetry is the existence of a current generator, $g$, with the property \cite{PhysRevX.6.041065,doyon2021free}
\begin{equation}
    \frac{\p g}{\p\beta^i}=\langle j_i\rangle_\rho.
\end{equation}
Note that $g$ is analogous to the free energy, in the sense that $\langle q_i\rangle=\p f/\p\beta^i$. In fact, on the basis on this analogy,
the form of  $g$ can be guessed such that the correct average currents  for integrable systems
are obtained \cite{PhysRevX.6.041065}.
A similar symmetry actually also holds for generalized currents $J_{jm}(x)$ which satisfy
\begin{equation}\label{generalizedcont}
    \ii[Q_m,q_n(x)]=j_{nm}(x)-j_{nm}(x+1).
\end{equation}
Note that in some literature the generalized currrents are defined with the opposite order of indices. Following the same arguments, it is then a simple matter to observe that the generalized charge-current susceptibility matrix
\begin{equation}
    B_{ijk}=\sum_{x\in\mathcal{
    S}} \langle q_i(x,0)j_{jk}(0,0)\rangle^\mathrm{c}_\rho
\end{equation}
satisfies the symmetry $B_{ijk}=B_{jik}$  \cite{10.21468/SciPostPhys.9.3.040,doyon2021free}. As before the symmetry allows us to define the generalized current generator $g_m$ by $\langle j_{nm}\rangle_\rho=\p g_m/\p\beta^n$. Curiously, the generators cannot be completely arbitrary, and are subject to the constraint $\sum_i\beta^ig_i=G$, where $G$ is some constant which vanishes for parity symmetric systems. This relation was discovered in \cite{doyon2021free} and dubbed Euler-scale KMS relation.

\subsection{Boost operators}
Usually, boost operators can be found in Galilean or Lorentz invariant systems, for which the operators literally boost the frame, equivalently, shift the rapidity as $\theta\mapsto \theta+\eta$. This property can be checked easily by recalling the algebra satisfied by the boost operator, which is simply the Poincar\'e algebra in (1+1)-dimensional relativistic QFTs,
\begin{equation}\label{poincare}
     [H,P]=0,\quad [B,H]=\ii P,\quad [B,P]=\ii H,
 \end{equation}
 where $B$ denotes the boost operator. The algebra immediately implies $e^{\ii B\eta}Pe^{-\ii B\eta}=(\cosh\eta) H-(\sinh\eta )P$, which in turn yields
 \begin{equation}
     Pe^{\ii B\eta}|\theta\rangle=m\sinh(\theta+\eta)e^{\ii B\eta}|\theta\rangle.
 \end{equation}
 Therefore the boosted state $e^{\ii B\eta}|\theta\rangle$ can be identified with another one-particle state with the boosted rapidity $\theta+\eta$. Such a property of the boost operator hints at the possibility that even in lattice systems without momentum conservation, one might define a boost operator provided that the constituent particles have stable rapidity. In other words, as in the continuum case, the boost operator acts on states by shifting rapidities. It is then natural to guess that integrable lattice models have a boost operator with the following property
\begin{equation}\label{boost1}
    e^{\ii \eta B}T(\lambda)e^{-\ii\eta B}=T(\lambda+\eta),
\end{equation}
where $T(\lambda)$ is the row-to-row transfer matrix of the model generating a tower of conserved charges. Such a boost operator does in fact exist in a certain class of integrable spin chains \cite{10.1143/PTP.69.431,THACKER1986348,PhysRevLett.58.1395}. To be more precise, the hamiltonian of the XYZ spin-1/2 chain (hence including the XXZ spin chain) with periodic boundary conditions reads $H=\sum_{j\in\mathcal{S}} h(j)$, where
\begin{equation}
   h(j)=-\tfrac{1}{2}\big(J_xS^x_jS^x_{j+1}+J_yS^y_jS^y_{j+1}+J_zS^z_jS^z_{j+1}\big).
\end{equation}
The boost operator of this model is known explicitly and given by $B=\sum_{x\in\mathcal{S}}xh(x)$ \cite{10.1143/PTP.69.431,THACKER1986348,PhysRevLett.58.1395}. Defining the conserved charges in a usual fashion as derivatives of the transfer operator, $ Q_n=\left.\frac{\dd^n}{\dd \lambda^n}\log T(\lambda)\right|_{\lambda=0}$, one can recast \eqref{boost1} into the boost algebra
\begin{equation}\label{boost}
     [B,Q_n]=\ii Q_{n+1}.
 \end{equation}
Notice that the boost algebra \eqref{boost} can be thought of as a lattice analogue of the Poincar\'e algebra. Indeed, in the relativistic continuum limit, the leading contribution to $Q_n$ arises from either $P$ (if $n$ is odd) or $H$ (if $n$ is even), since other terms are less local compared to these. Therefore 
the sine-Gordon/Thirring model, as continuum limit of the XYZ model, satisfies \eqref{poincare}.
 The availability of a boost operator has a significant implications. To see this, recall the continuity equation in spin chains
 \begin{equation}\label{latticecons}
     \ii[H,q_n(x)]=j_n(x)-j_n(x+1).
 \end{equation}
 Multiplying by $x$ on both sides and summing over $x\in\mathcal{S}$, one obtains
 \begin{equation}\label{boostcont}
     \ii[H,B_n]=\sum_{x\in\mathcal{S}}j_n(x)-Lj_n(1),
 \end{equation}
 where periodicity $j_n(1)=j_n(L+1)$ is used and $B_n=\sum_{x\in\mathcal{S}}xq_n(x)$ is the generalized boost operator (note that $B_1=B$, since $Q_1=H$). The second term on the right hand side is of no importance and can always be removed by redefining $j_n(x)\mapsto j_n(x)-j_n(1)$, which still satisfies the continuity equation \eqref{latticecons}, yielding $\ii[H,B_n]=J_n$ with $J_n=\sum_{x\in\mathcal{S}}j_n(x)$. Combining \eqref{boost} and \eqref{boostcont}, we arrive at the crucial observation 
 \begin{equation}\label{q3j2}
     Q_2=J_1,
 \end{equation}
where in the XXZ chain we label the charges as $Q_0=S_z, Q_1=H,\dots$\,.
This means that the current $J_1$ is one of the conserved charges, a property referred to as self-conserved. Apparently the above construction of a self-conserved current suggests that $J_1$ is the only self-conserved current under the Hamiltonian flow. In fact Eq. \eqref{q3j2} is readily generalized  to flows generated by higher spin conserved charges. The continuity equation \eqref{generalizedcont} yields $\ii[Q_m,B_n]=J_{nm}$, from which one infers that, for each $m$, there is a self-conserved current as $J_{1m}=Q_{m+1}$.

The above discussion deals with a specific model. One might then wonder if a boost operator as in \eqref{boost1} also exists in other known integrable spin chains. This turns out to be non-trivial question, since there are counter examples such as the Fermi-Hubbard model (FHM). The lack of the boost operator then indicates that there is no self-conserved current, which is in accordance with the nonvanishing energy diffusion in FHM, as observed in \cite{PhysRevLett.117.116401}. Algebraically, FHM has a structure very different from the XYZ chain, since the $R$-matrix of FHM is a solution of the Yang-Baxter equation which is not invariant under the boost $B$ \cite{PhysRevLett.86.5096}.

We are now in a position to spell out the identity which is central for establishing the collision rate ansatz. Eq. \eqref{q3j2} implies $C_{i2}=B_{i1}$ and, by the symmetry of the $B$ matrix, $C_{2i}=B_{1i}$ which can also be written as \cite{10.21468/SciPostPhys.9.3.040}
\begin{equation}\label{boostidentity1}
    \frac{\p}{\p\beta^2}\langle q_i\rangle= \frac{\p}{\p\beta^1}\langle j_i\rangle.
\end{equation}
The left hand side is a charge GGE susceptibility, which can be accessed through the TBA formalism,
while on the right the GGE averaged current appears. A similar relation holds whenever the model in question admits a self-conserved current, whether the system is integrable or not. 
The identity \eqref{boostidentity1} can be straightforwardly extended to the generalized current, which reads \cite{10.21468/SciPostPhys.9.3.040}
\begin{equation}
    \frac{\p}{\p\beta^{j+1}}\langle q_i\rangle=\frac{\p}{\p\beta^1}\langle j_{ij}\rangle.
\end{equation}

 \subsection{Establishing the rate ansatz}
  The proof requires only a single self-conserved current and in addition relies on the TBA formalism. To be concrete, we thus exemplify the method by working out the relativistic case of single particle species with diagonal scattering. In relativistic systems,  
  the energy current equals the momentum, i.e. $J_0=Q_1$ according to relativistic labelling conventions.  Therefore in this case the identity \eqref{boostidentity1} becomes
 \begin{equation}\label{boostidentity2}
     \frac{\p}{\p\beta^1}\langle q_i\rangle=\frac{\p}{\p\beta^0}\langle j_i\rangle.
 \end{equation}
 Progress is achieved by plugging in the identity \eqref{current}. But, at this point only linearity is used with some yet unknown function 
 $\bar{v}(\theta)$. Also inserted is the known TBA expression of the charge density average \eqref{chargeaverage}. Then
\begin{equation}\label{allj}
    \int_\mathbb{R}\dd\theta h_i(\theta)\p_{\beta^0}\big(\rho(\theta)\Bar{v}(\theta)\big)=\int_\mathbb{R}\dd\theta h_i(\theta)\p_{\beta^1}\rho(\theta).
\end{equation}
The goal is to show that $\bar{v}(\theta) = v^\mathrm{eff}(\theta)$.
Since \eqref{allj} is satisfied for all $i$, by completeness one concludes the pointwise identity
\begin{equation}\label{identity2}
    \p_{\beta^0}\big(\rho\Bar{v}\big) = \p_{\beta^1}\rho.
\end{equation}
 Let us first notice the obvious relations 
\begin{equation}\label{4.21}
 \p_{\beta^0}n = - n(1-n) (p')^\mathrm{dr}, \quad \p_{\mu_1}n = - n(1-n) (E')^\mathrm{dr}, 
\end{equation}
where  $p'(\theta)=E(\theta)$ and $E'(\theta)=p(\theta)$ has been used. This also implies
\begin{equation}\label{4.22}
\p_{\beta^1} (p')^\mathrm{dr} =  \p_{\beta^0}(E')^\mathrm{dr}
\end{equation}
with the dressing transformation defined in \eqref{dressing}. Hence
\begin{align}
\p_{\beta^0}(\rho v^\mathrm{eff})&=\frac{1}{2\pi}\left((E')^\mathrm{dr} \p_{\beta^0}n + n  \p_{\beta^0} (E')^\mathrm{dr}\right) \nonumber\\
 &= \frac{1}{2\pi}\left((p')^\mathrm{dr} \p_{\beta^1} n+ n \p_{\beta^1} (p')^\mathrm{dr}\right)\n
 &=  \frac{1}{2\pi}\p_{\beta^1}( (p')^\mathrm{dr}n) = \p_{\beta^1}\rho.
\end{align}
Therefore combining with \eqref{identity2}, we end up with $\p_{\beta^0}\big(\rho(\Bar{v}-v^\eff)\big)=0$, implying that $\rho(\Bar{v}-v^\eff)$ is independent of $\beta^0$. 
From the TBA equation \eqref{gge} one infers that the pseudo-energy $\varepsilon(\theta)\simeq \beta^0h_0(\theta)$ for $\beta^0\to\infty$ because $h_0(\theta)=m\cosh\theta>0$. Thus 
$n(\theta)=1/(1+e^{\varepsilon(\theta)})\to0$ and 
consequently $\rho^\tot(\theta)$ is uniformly bounded in $\beta^0$, implying $\rho$ to vanish. Assuming that $\bar{v}$ is also locally bounded for large
 $\beta^0$, as point-wise identity  $\rho\bar{v}\to0$ in the limit $\beta^0\to\infty$.  
 The free constant must be zero, thereby verifying the collision rate ansatz 
\begin{equation}
\Bar{v}=v^\eff. 
\end{equation}
 
 As mentioned before, the Fermi-Hubbard model has no self-conserved current and our argument fails. 
 Progress might be achieved through the suitably generalized boost operator of the Fermi-Hubbard model \cite{PhysRevLett.86.5096}.

\section{Summary and Outlook}
To a considerable extent generalized hydrodynamics stands on its own feet. 
The required microscopic input is rather minimal. One has to know the two-particle scattering shift,
mostly a simple task, and a free energy functional linked with the TBA formalism. The free energy 
functional also carries the distinction between bosonic, fermionic, and classical. With such an input
the hydrodynamic equations can be written. Their solution is a separate, mostly numerical issue.
Of great current interest is extending GHD to include integrability breaking terms,
as external potentials and small non-integrable interactions, for example \cite{durnin2021nonequilibrium,PhysRevLett.125.240604,PhysRevB.101.180302}.
GHD has developed into a powerful tool for our understanding of the dynamics of integrable many-body systems on large space-time scales.

From a theoretical perspective, one would like to start from a well-defined microscopic model
and then ensure the validity of GHD.  This turned out to be an ambitious, still ongoing program. 
Our review covers form factor expansions, mostly in the context of QFTs with diagonal scattering
and the Lieb-Liniger model as their non-relativistic limit. Transverse Ising and XYZ spin chains are also discussed. Form factors are a successful method in computing the effective velocity
as a nonlinear functional of the root density. An alternative approach is to use the symmetry
of the GGE charge-current correlator. Its connection to the boost algebra is explained.
Further aspects of the microscopic program are covered in companion reviews.

The hydrodynamic challenge has triggered novel developments in the form factor formalism.
But a microscopic derivation of the Drude weights as well as the diffusion matrix using the standard form factors are still outstanding tasks. Note that some hydrodynamic correlators were obtained along this line recently \cite{Vu_2020}. The form factor formalism may also prove capable of obtaining beyond-GHD regime corrections, including diffusive and quantum corrections, which can be compared to other current approaches \cite{de2019diffusion,ruggiero2020quantum,gopalakrishnan2018hydrodynamics}.

Strikingly, GHD  claims to apply to \textit{all} integrable many-body systems with quasi-local conservation laws. Best understood are quantum models solvable by Bethe, resp. asymptotic Bethe ansatz. An interesting model outside of this class is the Haldane-Shastry chain, which is not a Bethe-solvable model due to its non-local interactions \cite{PhysRevLett.66.1529}. Although the conservation laws of the model are constructed in an unusual way \cite{Bernard_1993}, integrability still leads to the existence of stable quasi-particles. Therefore it is tempting to expect that the hydrodynamic regime of the Haldane-Shastry chain is still captured by GHD.

A largely white area on the map are classical nonlinear wave equations, as the nonlinear Schr\"{o}dinger 
equation, the Korteweg de Vries equation, the classical sinh-Gordon equation, and others. Such models are ultraviolet divergent
and as a consequence the whole program of GGEs is somewhat pending. Nevertheless 
the hydrodynamic regime has been investigated \cite{PhysRevLett.95.204101,Carbone_2016,10.21468/SciPostPhys.4.6.045}. GHD has been studied for ultra-discrete wave equations, in particular the box-ball model \cite{Kuniba2020,Croydon2020}. Calogero-Moser models are another class of widely studied integrable classical systems
 \cite{Calogero2001}. The model closest to our review are $N$ classical point particles on a ring
interacting through the periodized $(\sinh(x))^{-2}$ pair potential. This system allows for a Lax matrix,
whose eigenvalues determine the quasi-local conserved charges. The quasi-local currents 
can also be expressed through the Lax matrix. The particle current is the momentum, hence conserved.
But the first road block occurs at establishing the generalized free energy functional. Classical models remain as a wide arena for quantitative tests of generalized hydrodynamics.

\section{Acknowledgements}
We are grateful to Alvise Bastianello, Jean-Sebasti\'en Caux, Jacopo De Nardis, Benjamin Doyon, Oleksandr Gamayun, Enej Ilievski, Marton Kormos,  Mi\l osz Panfil, Bal\'azs Pozsgay, Tomohiro Sasamoto, Dirk Schuricht, Dinh-Long Vu for discussions on topics that are presented in this review. We thank the editors for their constructive comments, which substantially improved the presentation of the review.

\bibliography{thesis}{}
\bibliographystyle{hieeetr.bst}

\end{document}